\documentclass{elsart}
\usepackage{graphicx,natbib,amssymb}
\bibpunct{(}{)}{;}{a}{}{,}

\def\fun#1#2{\lower3.6pt\vbox{\baselineskip0pt\lineskip.9pt
  \ialign{$\mathsurround=0pt#1\hfil##\hfil$\crcr#2\crcr\sim\crcr}}}
\def\lap{\mathrel{\mathpalette\fun <}}
\def\gap{\mathrel{\mathpalette\fun >}}

\newcommand{\kk}{\ensuremath{\,\mathrm{K}}}
\newcommand{\mm}{\ensuremath{\,\mathrm{M}}}

\newcommand{\GHz}{\ensuremath{\,\mathrm{GHz}}}
\newcommand{\GFlops}{\ensuremath{\,\mathrm{GFlops}}}
\newcommand{\TFlops}{\ensuremath{\,\mathrm{TFlops}}}
\newcommand{\Gbits}{\ensuremath{\,\mathrm{Gbit}\,\mathrm{s}^{-1}}}
\newcommand{\second}{\ensuremath{\,\mathrm{s}}}

\newcommand{\tmin}{\ensuremath{t_\mathrm{min}}}
\newcommand\ths{{T_{\rm host}}} 
\newcommand\tgr{{T_{\rm grape}}} 
\newcommand\tcm{{T_{\rm comm}}} 
\newcommand\tpr{{T_{\rm pred}}} 
\newcommand\tcr{{T_{\rm corr}}} 
\newcommand\tpp{{T_{\rm pipe}}}
\newcommand\tmpi{{T_{\rm MPI}}}

\newcommand{\myddot}[1]{\mathbf{#1}^{(2)}}
\newcommand{\mydddot}[1]{\mathbf{#1}^{(3)}}

\begin{document}

\begin{frontmatter}

%\authorrunninghead{Harfst et al.}
%\titlerunninghead{Special-Purpose Supercomputers}

%\setcounter{page}{261} %% This command is optional. 

%% <<== End of commands to be entered at Academic Press

\title{Performance Analysis of Direct $N$-Body Algorithms on Special-Purpose Supercomputers}
\author[RIT]{Stefan Harfst},
\ead{harfst@astro.rit.edu}
\author[AIAP,RSDN]{Alessia Gualandris},
\author[RIT]{David Merritt},
\author[ARI,RSDN]{Rainer Spurzem},
\author[AIAP,RSDN]{Simon Portegies Zwart},
\author[ARI,KIEV,RSDN]{Peter Berczik}

\address[RIT]{Department of Physics and Astronomy, Rochester Institute of Technology, Rochester, NY 14623} 
\address[AIAP]{Astronomical Institute {\it Anton Pannekoek} and Section 
Computational Science, University of Amsterdam, The Netherlands}
\address[ARI]{Astronomisches Rechen-Institut, Zentrum f\"ur Astronomie,
Universit\"at Heidelberg, 
Heidelberg, Germany}
\address[KIEV]{Main Astronomical Observatory, National Academy of Science, Kiev, Ukraine, 03680}
\address[RSDN]{The Rhine Stellar Dynamical Network}

\begin{abstract}
Direct-summation $N$-body algorithms compute the gravitational
interaction between stars in an exact way and have a
computational complexity of $\mathcal{O}(N^2)$.  Performance can be greatly
enhanced via the use of special-purpose accelerator boards like
the GRAPE-6A. However the memory of the GRAPE boards is limited.  
Here, we present a performance analysis of direct $N$-body codes on two
parallel supercomputers that incorporate special-purpose boards,
allowing as many as four million particles to be integrated.
Both computers employ high-speed, Infiniband interconnects to
minimize communication overhead, which can otherwise become
significant due to the small number of ``active'' particles at
each time step.  We find that the computation time scales well
with processor number; for $2\times 10^6$ particles, efficiencies
greater than 50\% and speeds in excess of $2\TFlops$ are reached.
\end{abstract}



\end{frontmatter}

\section{Introduction}

Numerical algorithms for solving the gravitational $N$-body problem
\citep{aarseth-03} have evolved along two basic lines in recent years.  
Direct-summation
codes compute the complete set of $N^2$ interparticle forces at each
time step; these codes are designed for systems in which 
the finite-$N$ graininess of the potential is important or in
which binary- or multiple-star systems form,
and until recently, were limited by their $\mathcal{O}(N^2)$ scaling to moderate 
($N\lap 10^5$) particle
numbers.  The best-known examples are the {\tt NBODY} series of codes
introduced by \cite{aarseth-99} and the {\tt Starlab}
environment developed by McMillan, Hut, and collaborators
(e.g. \cite{STARLAB}).

A second class of $N$-body algorithms replace the direct summation of
forces from distant particles by an approximation scheme.  Examples
are the Barnes-Hut tree code \citep{bh-86}, which reduces the number of
force calculations by subdividing particles into an oct-tree, and fast
multipole algorithms which represent the large-scale potential via a
truncated basis-set expansion \citep{vavg-77,gr-87}, or on a grid
\citep{mp-68,ee-81}.  These algorithms have a milder, $\mathcal{O}(N\log N)$ or even
$\mathcal{O}(N)$ scaling for the force calculations and can handle much larger
particle numbers, although their accuracy can be substantially lower
than that of the direct-summation codes \citep{spurzem-99}.
The efficiency of both sorts of algorithm can be considerably
increased by the use of individual time steps for advancing particle
positions \citep{aarseth-03}.

The $N$-body problem is particularly challenging in the case of
dense stellar systems like galactic nuclei, which may contain
single or multiple, supermassive black holes in addition
to stars \citep{FF05,living}.  
Particle advancement must be very accurate
for trajectories that pass near the black hole(s).  
In addition, galactic nuclei are often ``collisional'' in the sense that
gravitational encounters (i.e. scattering) can redistribute energy
between stars on time scales less than the age of the universe
\citep{merritt-06}.
Simulating the evolution of a collisional nucleus is probably
not feasible using tree or grid algorithms due to their limited accuracies.
While alternative, highly-efficient algorithms based on
the Fokker-Planck or fluid formalisms have been widely
used to model galactic nuclei \citep{ls-91,fb-01},
these methods can not deal with systems that are far from
dynamical equilibrium or that fail to respect spatial symmetries.

A natural way to increase both the speed and particle number in an
$N$-body simulation is to parallelize \citep{dubinski-96,pc-97}.  
Parallelization on
general-purpose supercomputers is difficult, however, because the
calculation cost is often dominated by a small number of particles in
a single dense region, e.g. the nucleus of a galaxy.
Communication latency becomes the bottleneck: the time to communicate
particle positions between processors can exceed the time spent
computing the forces.  The best such schemes use systolic algorithms
(in which the particles are rhythmically passed around a ring of
processors) coupled with non-blocking communication between the
processors to reduce the latency \citep{makino-02,dorband-03}.

A major breakthrough in direct-summation $N$-body simulations came in
the late 1990s with the development of the GRAPE series of
special-purpose computers \citep{grapebook}, which achieve spectacular
speedups by implementing the entire force calculation in hardware and
placing many force pipelines on a single chip.  The GRAPE-6, in its
standard implementation (32 chips, 192 pipelines), can achieve
sustained speeds of about 1 Tflops at a cost of just $\sim\$50$K.
In a standard setup, the GRAPE-6 is attached to a single host workstation,
in much the same way that a floating-point or graphics accelerator
card is used.  Advancement of particle positions [$\mathcal{O}(N)$] is carried
out on the host computer, while interparticle forces [$\mathcal{O}(N^2)$] are
computed on the GRAPE. More recently, ``mini-GRAPEs'' (GRAPE-6A)
\citep{FMK05} have become available, which are designed to be
incorporated into the nodes of a parallel computer.
The mini-GRAPEs place four processor chips
on a single PCI card and delivering a theoretical peak performance of
$\sim 131$ Gflops for systems of up to 128k particles,
at a cost of $\sim\$6$K.
By incorporating mini-GRAPEs into a cluster,
both large ($\gap 10^6$) particle numbers and high
($\gap 1\TFlops$) speeds can in principle be achieved.

In this paper, we describe the performance of direct-summation
$N$-body algorithms on two computer clusters that incorporate GRAPE
hardware.  
\S 2 describes the hardware implementations.
The parallel $N$-body code and its implementation on the
GRAPEs is described in \S 3.
\S 4 presents the results of performance tests using
realistic galaxy models, and \S 5 describes a theoretical
performance model that reproduces the observed performance
and which can be used to predict the performance of similar
codes on different clusters.
\S 6 summarizes our results and discusses directions for future
work.

%__________________________________________________________________________
%
%  Section: Hardware
%
\section{Hardware}

\subsection{The GRAPE technology}

\begin{figure}[t]
\begin{center}
\resizebox{\hsize}{!}{\includegraphics{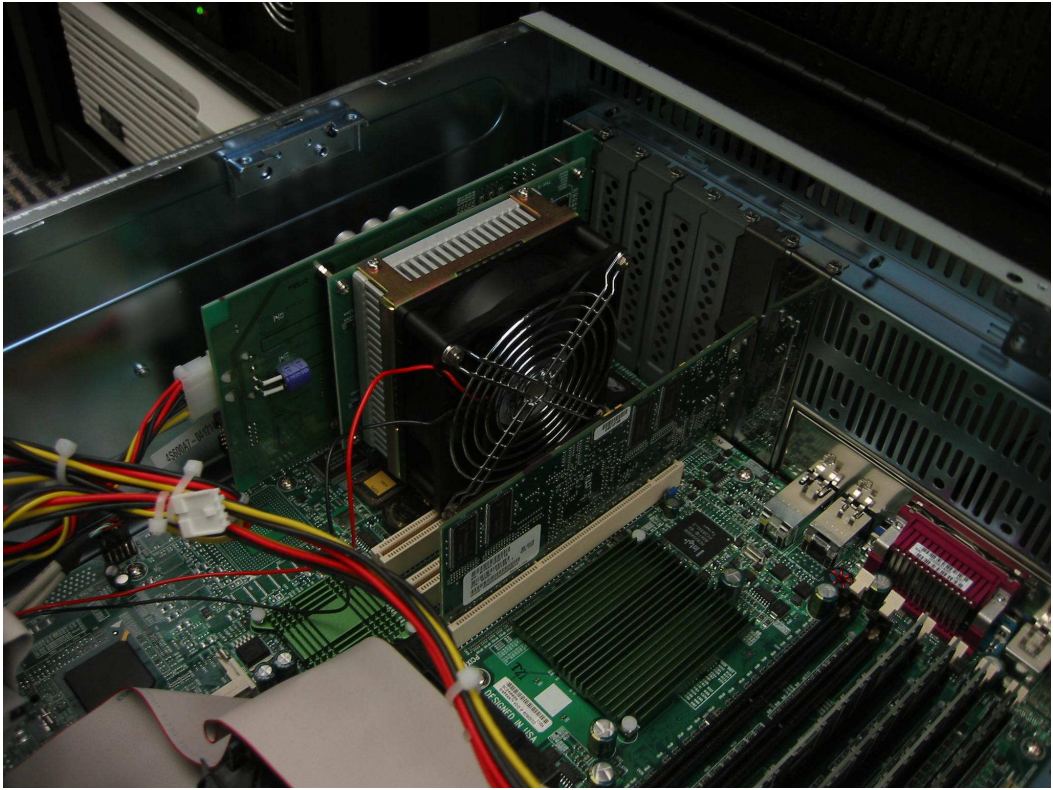}}
\end{center}
\caption{Interior of a node showing a GRAPE-6A card 
(note the large black fan) and an Infiniband card. } 
\label{fig_nodeint}
\end{figure}

The GRAPE-6A board (Fig.~\ref{fig_nodeint}) is a standard PCI short
card on which a processor, an interface unit, and a power supply are
integrated.  The processor is a module consisting of four GRAPE-6
processor chips, eight SSRAM chips and one FPGA chip.  The processor
chips each contain six force calculation pipelines, a predictor
pipeline, a memory interface, a control unit, and I/O ports
\citep{MFK03}.  The SSRAM chips store the particle data.  The four
GRAPE chips can calculate forces, their time derivatives and the
scalar gravitational potential simultaneously on a maximum of 48
particles at a time; this limit is set by the number of pipelines (six
force calculation pipelines each of which serves as eight virtual
multiple pipelines). There is also a facility to calculate neighbor
lists from predefined neighbor search radii; this feature is not used 
in the algorithms presented below. The forces computed by the processor chips
are summed in an FPGA chip and sent to the host computer.  A maximum
of 131\,072 ($2^{17}$) particles can be stored in the GRAPE-6A memory.
The peak speed of the GRAPE-6A is 131.3$\GFlops$ (when computing
forces and their derivatives) and 87.5$\GFlops$ (forces only),
assuming 57 floating-point operations per force calculation
\citep{FMK05}.  The interface to the host computer is via a standard 32
bit/33 MHz PCI bus.  The FPGA chip (Altera EP1K100FC256) realizes a
4-input, 1-output reduction when transferring data from the GRAPE-6
processor chip to host computer.  The complete GRAPE-6A unit
(Fig.~\ref{fig_nodeint}) is roughly 11 cm $\times$ 19 cm $\times$ 7 cm
in size. 5.8 cm of the height are taken up by a rather bulky
combination of cooling body and fan, which may block other slots on
the main board. Possible ways to deal with this include the use
of even taller boxes for the nodes (e.g. 5U) together with a
PCI riser of up to 6 cm, which would allow the use of slots for
interface cards beneath the GRAPE fan; or the adoption of the more 
recent, flatter designs such as that of the GRAPE6-BL series. 
The reader interested in more technical details should seek advice from 
the GRAPE ({\tt http://astrogrape.org})
and Hamamatsu Metrix ({\tt http:/www.metrix.co.jp})
websites.

%---------------------------------------------------------------------
%   Figure: Picture of a node's interior
%   ------------------------------------
%------------------- end of figure NT1 ---------------------------------

\subsection{The GRAPE cluster}

\begin{figure}[t]
\begin{center}
\resizebox{\hsize}{!}{\includegraphics{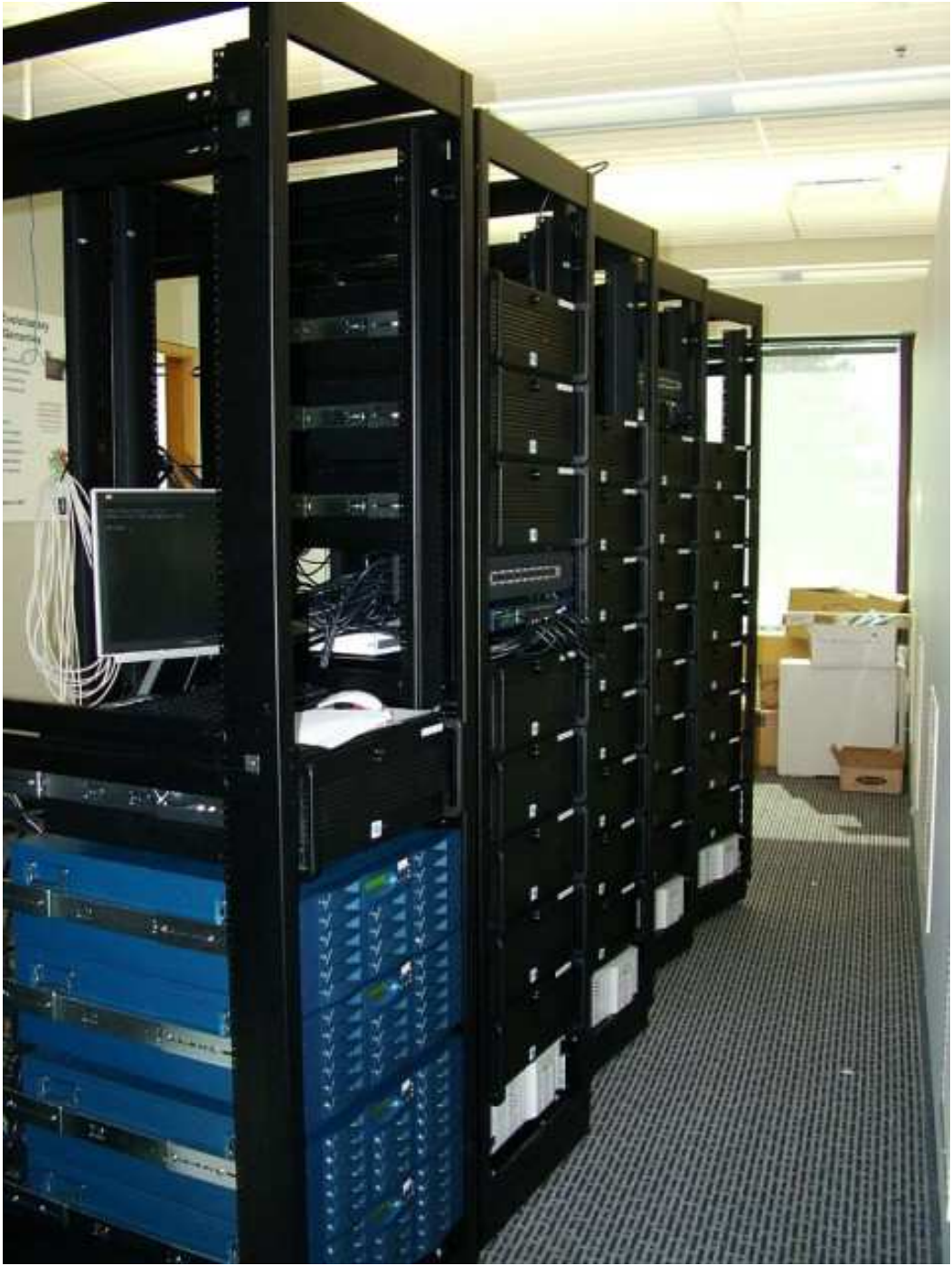}}
\end{center}
\caption{gravitySimulator, the GRAPE cluster at RIT. 
The head node and the 14\,Tbyte raid array are visible 
on the first rack. 
The other four racks hold a total of 32 compute nodes,
 each equipped with a GRAPE-6A card.} 
\label{fig_clusterpic}
\end{figure}

A computer cluster incorporating GRAPE-6A boards became fully
operational at the Rochester Institute of Technology (RIT) in
February 2005 (Fig.~\ref{fig_clusterpic}).  This cluster, named
``gravitySimulator,'' consists of 32 compute nodes plus one head
node each containing dual $3\GHz$-Xeon processors.  In addition
to a standard Gbit-ethernet, the nodes are connected via a
low-latency Infiniband network with a transfer rate of
$10\Gbits$.  The typical latency for an Infiniband network is of
the order of $10^{-6}\second$, or a factor $\sim 100$ better then
the Gbit-Ethernet \citep{liu-03}.  A total of 14\,TByte of disk
space is available on a level 5 RAID array.  The disk space is
equivalent to $2.5\times 10^5$ $N$-body data sets each with
$10^6$ particles.  The disks are accessed via a fast Ultra320
SCSI host adapter from the head node or via NFS from the compute
nodes, which in addition are each fitted with a 80 Gbyte hard
disk.  Each compute node also contains a GRAPE-6A PCI card
(Fig.~\ref{fig_nodeint}).  The total, theoretical peak
performance is approximately $4\TFlops$ if the GRAPE boards
are fully utilized.
Total cost was roughly \$0.45M, roughly 1/2 of which was
used to purchase the GRAPE boards.

%---------------------------------------------------------------------
%   Figure: Picture of the RIT cluster
%   ----------------------------------
%------------------- end of figure -----------------------------------

Some special considerations were required in order to incorporate
the GRAPE cards into the cluster.  Since our GRAPE-6A's use the
relatively old PCI interface standard (32\,bit/33\,MHz), only one
motherboard was found, the SuperMicro X5DPL-iGM, that could
accept both the GRAPE-6A and the Infiniband card. (A newer
version of the GRAPE-6A that uses the faster PCI-X technology is
now available.) The PC case itself has to be tall enough (4U) to
accept the GRAPE-6A card and must also allow good air flow for
cooling since the GRAPE card is a substantial heat source. The
cluster has a total power consumption of $17\,\mathrm{kW}$
when the GRAPEs are fully loaded.
Cluster cooling was achieved at minimal cost by redirecting the
air conditioning from a large room toward the air-intake side of
the cluster.  Temperatures measured in the PC case and at the two
CPUs remain below 30\,C and 50\,C, respectively.

A similar cluster, called ``GRACE'' (GRAPE + MPRACE), was recently
installed in the Astronomisches Rechen-institut (ARI) at the
University of Heidelberg.  There are two major differences
between the RIT and ARI clusters.  1) Each node of the ARI 
cluster incorporates a reconfigurable FPGA card (called ``MPRACE'') 
in addition to to the GRAPE board.
MPRACE is optimized to compute neighbor forces and other,
non-Newtonian forces between particles, in order to accelerate
calculations of molecular dynamics, smoothed-particle hydrodynamics,
etc. 
2) The newer main board SuperMicro X6DAE-G2 was used, 
which supports Pentium Xeon chips with 64 bit technology (EM64T) 
and the PCIe (PCI express) bus.
This made it possible to use dual-port Infiniband interconnects 
via the PCI express Infiniband x8 host interface card,
used in the x16 Infiniband slot of the board (it has another x4 Infiniband
slot which is reserved for the MPRACE-2 Infiniband card). 
As discussed below, the use of the PCIe bus substantially
reduces communication overhead.
Benchmark results presented below for the ARI cluster were
obtained from algorithms that do not access the FPGA cards.
%We think that the use
%of reconfigurable logic, presently a relatively new approach to
%high-performance computing and its power should increase
%dramatically in the future \citep{alfke-01,hamadaetal05,nakasatoetal06,
%spurzemetal06}.

\section{The parallel $N$-body code}
\label{sec:code}

We present here a new direct-summation code, called $\varphi$GRAPE, which
was optimized for perfomance on GRAPE clusters. 
The algorithm employs a Hermite integration scheme \citep{MA92} with 
hierarchical, commensurate block time steps.  
Although comparable in complexity to -- and inspired by -- Aarseth's 
serial NBODY1 code \citep{aarseth-99}, 
$\varphi$GRAPE was written ``from scratch'' for the RIT GRAPE
cluster. 
\footnote{The code will be publicly available; see
{\tt http://grapecluster.rit.edu/}.} 

\subsection{Integration scheme}

In addition to position ${\bf x}_i$, velocity ${\bf v}_i$,
acceleration ${\bf a}_i$, and time derivative of acceleration ${\bf
\dot{a}}_i$, each particle $i$ has its own time $t_i$ and time step
$\Delta t_i$.

Integration consists of the following steps:
\begin{enumerate}
\item
The initial time steps are calculated from
\begin{equation}
\Delta t_i = \eta_s\frac{|{\bf a}_{i}|}{|{\bf \dot{a}}_{i}|},
\end{equation}
where typically $\eta_s = 0.01$ gives sufficient accuracy.
\medskip
\item The system time $t$ is set to the minimum of all $t_i + \Delta t_i$ and
all particles $i$ that have  $t_i + \Delta t_i=t$ are selected as
active particles. Note that ``classical'' $N$-body codes \citep{aarseth-99,
aarseth-03} employ a sorted time-step list to select short time step particles
efficiently. Such sorting is abandoned in favor of a search over
all $N$ particles, which improves load balance in the parallel code
due to the more random distribution of short and long step particles.
\medskip
\item Positions and velocities at the new $t$ are predicted for all
particles using 
\begin{equation}
{\bf x}_{j,\mathrm{p}} = {\bf x}_{j,0} + (t-t_j) {\bf v}_{j,0} + 
\frac{(t-t_j)^2}{2}{\bf a}_{j,0} + \frac{(t-t_j)^3}{6}{\bf \dot{a}}_{j,0}
\end{equation}
and
\begin{equation}
{\bf v}_{j,\mathrm{p}} = {\bf v}_{j,0} +
(t-t_j){\bf a}_{j,0} + \frac{(t-t_j)^2}{2}{\bf \dot{a}}_{j,0}.
\end{equation}
Here, the second subscript denotes a value given either at the
beginning (0) or the end (1) of the current time step. All quantities
used in the predictor can be calculated directly, i.e. no memory of a
previous time step is required.  
\medskip
\item Acceleration and its time derivative are updated for active
particles only according to
\begin{equation}
{\bf a}_{i,1} = \sum_{j\ne i} Gm_j\frac{{\bf
r}_{ij}}{(r_{ij}^2+\epsilon^2)^{(3/2)}}
\end{equation}
and
\begin{equation}
{\bf \dot{a}}_{i,1} = \sum_{j\ne i} Gm_j\left[\frac{{\bf
v}_{ij}}{(r_{ij}^2+\epsilon^2)^{(3/2)}} + \frac{3({\bf
v}_{ij}\cdot{\bf r}_{ij}){\bf r}_{ij}}{(r_{ij}^2+\epsilon^2)^{(5/2)}}\right],
\end{equation}
where
\begin{equation}
{\bf r}_{ij} = {\bf x}_{j,p}-{\bf x}_{i,p},
\end{equation}
\begin{equation}
{\bf v}_{ij} = {\bf v}_{j,p}-{\bf v}_{i,p},
\end{equation}
and $\epsilon$ is the softening parameter.
\medskip
\item Positions and velocities of active particles are corrected
using 
\begin{equation}
{\bf x}_{i,1} =  {\bf x}_{i,\mathrm{p}} 
               + \frac{\Delta t_i^4}{24}\myddot{a}_{i,0} 
               + \frac{\Delta t_i^5}{120}\mydddot{a}_{i,0} 
\end{equation}
and
\begin{equation}
{\bf v}_{i,1} =  {\bf v}_{i,\mathrm{p}} 
               + \frac{\Delta t_i^3}{6}\myddot{a}_{i,0} 
               + \frac{\Delta t_i^4}{24}\mydddot{a}_{i,0}, 
\end{equation}
where the second and third time derivatives of ${\bf a}$ are given by
\begin{equation}
\myddot{a}_{i,0} = \frac{-6\left({\bf a}_{i,0}-{\bf a}_{i,1}\right)
	                       - \Delta t_i 
                    \left(4{\bf \dot{a}}_{i,0}+2{\bf \dot{a}}_{i,1}\right)}
                           {\Delta t_i^2}
\end{equation}
\begin{equation}
\mydddot{a}_{i,0} = \frac{12\left({\bf a}_{i,0}-{\bf a}_{i,1}\right)
	                       +6 \Delta t_i 
                    \left({\bf \dot{a}}_{i,0}+{\bf \dot{a}}_{i,1}\right)}
                           {\Delta t_i^3}.
\end{equation}
\medskip
\item The times $t_i$ are updated and the new time steps
$\Delta t_i$ are determined.
Time steps are calculated using the standard formula \citep{A85}:
\begin{equation}
\Delta t_{i,1} = \sqrt{\eta\displaystyle\frac{|{\bf a}_{i,1}||
\myddot{a}_{i,1}| + |{\bf \dot{a}}_{i,1}|^2}
{|{\bf \dot{a}}_{i,1}||\mydddot{a}_{i,1}| + |\myddot{a}_{i,1}|^2}}.
\label{eq_timestep}
\end{equation}
The parameter $\eta$ controls the accuracy of the integration and is
typically set to $0.02$. 
The value of $\myddot{a}_{i,1}$ is calculated from 
\begin{equation}
\myddot{a}_{i,1} = \myddot{a}_{i,0} + \Delta t_{i,0}\mydddot{a}_{i,0}
\end{equation}
and $\mydddot{a}_{i,1}$ is set to $\mydddot{a}_{i,0}$.
\medskip
\item Repeat from step (2).
\end{enumerate}

A hierarchical commensurate
block time step scheme is necessary when the Hermite integrator
is used with the GRAPE (and is also efficient for parallelization and
vectorization; see
below and \citet{mcmillan-86}). 
Particles are grouped by replacing their time steps $\Delta
t_i$ with a block time step $\Delta t_{i,\mathrm{b}}=(1/2)^n$, where
$n$ is chosen according to
\begin{equation}
\left(\frac{1}{2}\right)^n \le \Delta t_i < \left(\frac{1}{2}\right)^{n-1}.
\end{equation} 
The commensurability is enforced by requiring that $t/\Delta t_i$ 
be an integer.
For numerical reason we also set a minimum time step $\Delta t_\mathrm{min}$,
where typically
\begin{equation}
\Delta t_\mathrm{min} = 2^{-23}.
\label{eq_tmin}
\end{equation}
The time steps of particles with $\Delta t_i < \Delta t_\mathrm{min}$
are set to this value. The minimum time step should be consistent with
the maximum acceleration defined by the softening parameter;
monitoring of the total energy can generally indicate whether this
condition is being violated.

\subsection{GRAPE implementation}
\label{sec_G6impl}

The GRAPE-6 and GRAPE-6A hardware has been designed to work with a
Hermite integration scheme and is therefore easily integrated into the
algorithm described in the previous section \citep[see][]{MFK03}. 
In detail, integration of particle positions using the GRAPE-6A
consists of the following steps:

\begin{enumerate}
\item \textbf{Initialize} the GRAPE and send particle data (positions, velocities,
etc.) to GRAPE memory. 
\\
\item \textbf{Compute} the next system time $t$ and select active particles on
the host (same as step 2 in previous section).
\\
\item \textbf{Predict} postions and velocities of active particles only and
send the predicted values together with the new system time $t$ to
GRAPE's force calculation pipeline.
\\
\item \textbf{Predict} positions and velocities for all other
particles on the GRAPE, and calculate forces and their time 
derivatives for active particles.
\\
\item \textbf{Retrieve} forces and their time derivatives from the
GRAPE and correct postions and velocities of active particles
on the host.
\\
\item \textbf{Compute} the new time steps and update the particle data
on the host of all active particles in the GRAPE memeory. 
\\
\item \textbf{Repeat} from step (2).
\end{enumerate}

\subsection{Parallelization}

Two basic schemes have been used to implement parallel force
computations for $\mathcal{O}(N^2)$ problems on general-purpose computers. 
The simplest case to consider is when all particles have the same fixed
time step.

\noindent {\it Replicated data algorithms (also called ``copy'' 
or ``broadcast'' algorithms).}
Each compute node has a copy of the whole system but
is assigned a specific subset of $N/p$ particles,
where $p$ is the number of processors.
At every step, each node computes the forces exerted by
all $N$ particles on its subset.
These particles are then advanced and their updated positions 
and velocities are sent to the other processors. 

\noindent {\it Systolic algorithms (also called ``ring'' algorithms).}
At the start of the integration, each node is permanently assigned a 
subset of $N/p$ particles.
At each step, these sub-arrays are shifted sequentially to the other
nodes where the partial forces are computed and stored.
After $p-1$ such shifts, all of the force pairs have been computed
and the particles are returned to their original nodes 
where their trajectories can be advanced.

\noindent Both the replicated and systolic algorithms exhibit an
$\mathcal{O}(N\log p)$ scaling in communication complexity and an
$\mathcal{O}(N^2)$ scaling in number of force computations.
The systolic algorithm makes more efficient use of memory;
however memory limitations are typically not restrictive
for $N\lap 10^6$ \citep{dorband-03,gual-05}.

The performance of parallel algorithms can be substantially
degraded however if the $N$-body system has a ``core-halo''
structure, i.e.  a dense central region surrounded by a
low-density envelope.  A galaxy containing a central black hole
is an extreme example.  Individual time steps
(Equation~\ref{eq_timestep}) are mandated in this case, and the
group size -- the number of active particles due to be advanced
at every time step -- can be much smaller than $N$; indeed it can
often be smaller than $p$.  In the latter case, the systolic
algorithm suffers since only a fraction of the nodes are active
at a given time.  Nonblocking communication is an effective way
to deal with this problem since it allows communication to be put
``in the background'' so that computing nodes can send/receive
data and calculate at the same time \citep{dorband-03}.

Adding the GRAPE hardware imposes additional constraints.
The GRAPE memory holds only $N_G\approx 10^5$ particles,
hence the copy algorithm becomes inefficient for large $N$.
The systolic algorithm is a natural alternative, allowing
a total of $N_G\times p$ particles to be stored in the 
collective GRAPE memories.
But a problem arises with the systolic algorithm
if the number of active particles
on any node is less than 48, since the time required by the
GRAPE to compute forces on one particle is the same as the time
to compute the forces on 48.

Our solution was to adopt a hybrid scheme. 
Nodes are initially assigned a subset of $N/p$ particles
as in the systolic algorithm, where $N/p \leq N_G$, ensuring
that all $N/p$ particles can be stored in the GRAPE
memory.
However, once the active particles on each node have been
identified, they are broadcast to all the other nodes,
thus minimizing the possibility that any one GRAPE will
be required to compute forces on less than 48 
particles.\footnote{\citet{FMK05} have adopted a similar scheme.}

In detail, our parallel algorithm works as follows:

\begin{enumerate}
\item \textbf{Distribute} the particle data to all nodes
such that each node receives $N/p$ particles.
\\
\item \textbf{Initialize} the GRAPE card on each node and 
send the local particle data to the GRAPE memories.
\\
\item \textbf{Compute} the minimum time step on each node and use {\tt
allreduce} to find the global minimum. 
\\
\item \textbf{Select} the active particles on each node
and predict their positions and velocities.
\\
\item \textbf{Collect} the particle data (including the predicted 
values) of all active particles onto all nodes using {\tt allgather}.
\\
\item \textbf{Compute} the partial forces on each node
for the global set of active particles using the GRAPE.
\\
\item \textbf{Retrieve} the local partial forces, which are summed
to get the total forces using {\tt allreduce}.
\\
\item \textbf{Correct} positions and velocities for the local
active particles on each node and update the GRAPE memory.
\\
\item \textbf{Repeat}  from step (3).
\end{enumerate}

\section{Performance tests}

We evaluated the performance of this algorithm on the two
GRAPE clusters described above.

\subsection{$N$-body models}

Initial conditions for the performance tests were produced by
generating Monte-Carlo positions and velocities from
self-consistent models of stellar systems.  Each of these models
is spherical and is completely described by a steady-state
phase-space distribution function $f(E)$ and its self-consistent
potential $\Psi(r)$, where $E=v^2/2+\Psi$ is the particle energy
and $r$ is the distance from the center.  The models were: a
\cite{P11} sphere; two \cite{K66} models with different
concentrations; and two \cite{D93} models with different central
density slopes.  The Plummer model has a low central
concentration and a finite central density; it does not
accurately represent any class of stellar system but is a common
test case.  King models are defined by a single dimensionless
parameter $W_0$ describing the central concentration (e.g. ratio
of central to mean density); we used $W_0 = 9$ and $W_0 = 12$
which are appropriate for globular star clusters
\citep{spitzer-87}.  Dehnen models have a divergent inner density
profile, $\rho\propto r^{-\gamma}$.  We took $\gamma = 0.5$ and
$\gamma = 1.5$, which correspond approximately to the inner
density profiles of bright and faint elliptical galaxies
respectively \citep{gebhardt-96}; in particular, the central
bulge of the Milky Way galaxy has $\rho\sim r^{-1.5}$
\citep{genzel-03}.

In what follows we adopt standard $N$-body units $G=M=-4E=1$, where
$G$ is the gravitational constant, $M$ the total mass and $E$ the
total energy of the system. In some of the models, the initial time
step for some particles was smaller than the minimum time step $\tmin$
set in Eq.~\ref{eq_tmin}. These models were then rescaled to change
the minimum time step to a large enough value. Since the rescaling
does not influence the performance results we will present all results
in the standard $N$-body units.

We realized each of the five models with 11 different particle
numbers, $N=2^k$, $k=[10,11,...,20]$, i.e. $N=[1\kk,
2\kk,...,1\mm]$.\footnote{Henceforth we use K to denote a factor of 
$2^{10}=1024$ and M to denote a factor of $2^{20}=1,048,576$.}  We also tested
Plummer models with $N=2\mm$ and $N=4\mm$; the latter value is the
maximum $N$ value allowed by filling the memory of all 32 GRAPE
cards. Thus, a total of 57 test models were used in the timing runs.

%---------------------------------------------------------------------
%   Figure: Energy conservation all models
%   ------------------------------------------
\begin{figure}[t]
\begin{center}
\resizebox{\hsize}{!}{\includegraphics[angle=90]{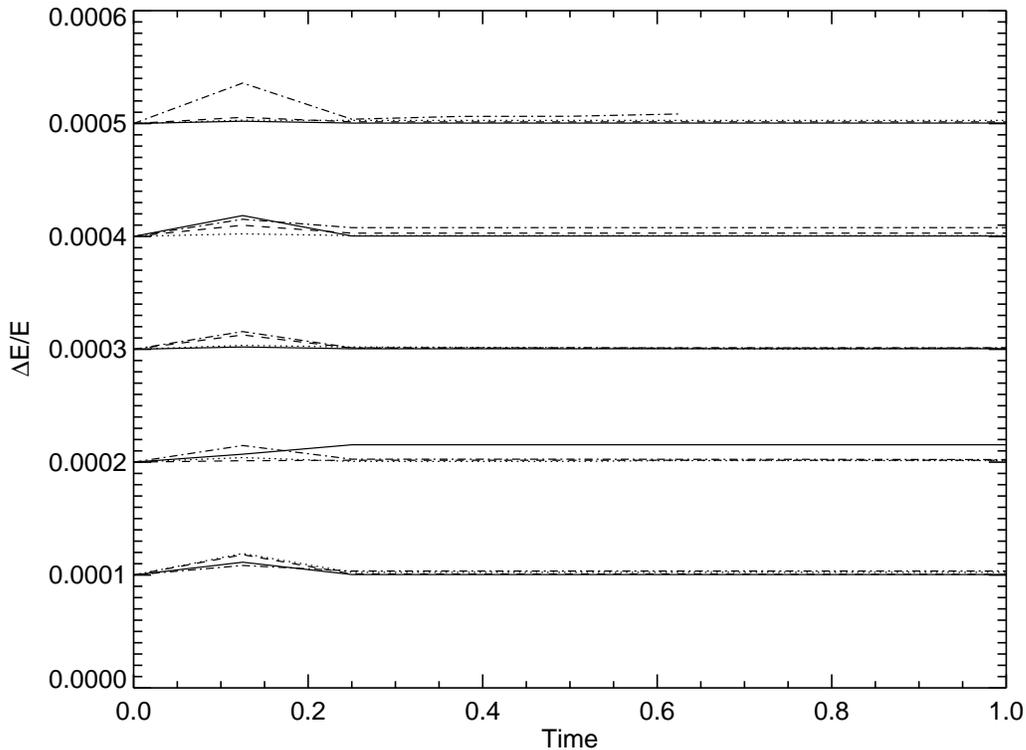}}
\end{center}
\caption{Energy conservation in each of the five test models,
for four $N$-values.
Relative errors have been shifted by multiples of $10^{-4}$ for
each model. From top to bottom, each group of four lines represents 
Plummer models; Dehnen modes with $\gamma=0.5$ and $\gamma=1.5$; and King
models with $W_0=9$ and $W_0=12$. The results for 
$N=4\kk$ (full line), $64\kk$ (dotted), $256\kk$ (dashed) and
$512\kk$ (dash-dotted) are shown in each group.} 
\label{fig_denergy}
\end{figure}
%------------------- end of figure -----------------------------------

%---------------------------------------------------------------------
%   Figure: Lagrange radii for the Plummer model
%   --------------------------------------------
\begin{figure}[t]
\begin{center}
\resizebox{\hsize}{!}{\includegraphics[angle=90]{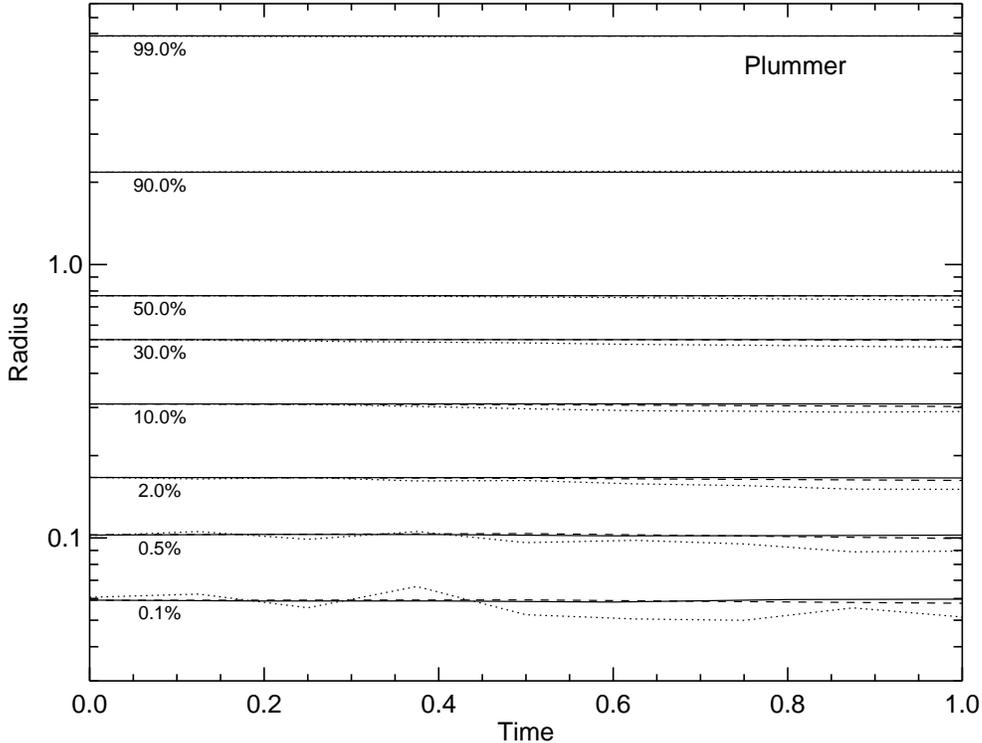}}
\end{center}
\caption{Lagrange radii for Dehnen models ($\gamma=1.5$) with
different $N$. The results
for 4\,096 (dotted), 131\,072 (dashed) and 1\,048\,576 particles (full
line) are shown. 
This is the densest (i.e. most centrally concentrated) of the
test models are represents appoximately density profile
of the Milky Way bulge.
Lagrange radii of all other tested models were more stable.} 
\label{fig_lagrange}
\end{figure}
%------------------- end of figure  -----------------------------------

Two-body relaxation, i.e. exchange of energy between particles due to
gravitational scattering, induces a slow change in the characteristics
of the models.  In order to minimize the effects of these changes on
the timing runs, we integrated the models for only one time unit. The
softening $\epsilon$ was set to zero for the Plummer models and to
$10^{-4}$ for the Dehnen and King models. The time step parameters
were $\eta_s=0.01$ and $\eta=0.02$.

Figs.~\ref{fig_denergy} and \ref{fig_lagrange} show the dependence on
time of the total energy, and the Lagrange radii, for the models.  In
all models, the maximum relative deviation in total energy is of the
order of $10^{-5}$ or less. The Lagrange radii (shown only for
Dehnen models with $\gamma=1.5$, the most centrally concentrated
of the models which we considered) show that the mass
profiles of all models remain practically unchanged. A noticable, but
small, change in the innermost region can be seen only for the
lowest particle numbers.

\subsection{Performance results}

%----------------------------------------------------------------------
%
%   Figure: Performance plots - Wallclock times
%   ===========================================
%
\begin{figure} 
   \resizebox{\hsize}{!}{  
   \begin{tabular}{cc}
   \includegraphics[width=\hsize,angle=90]{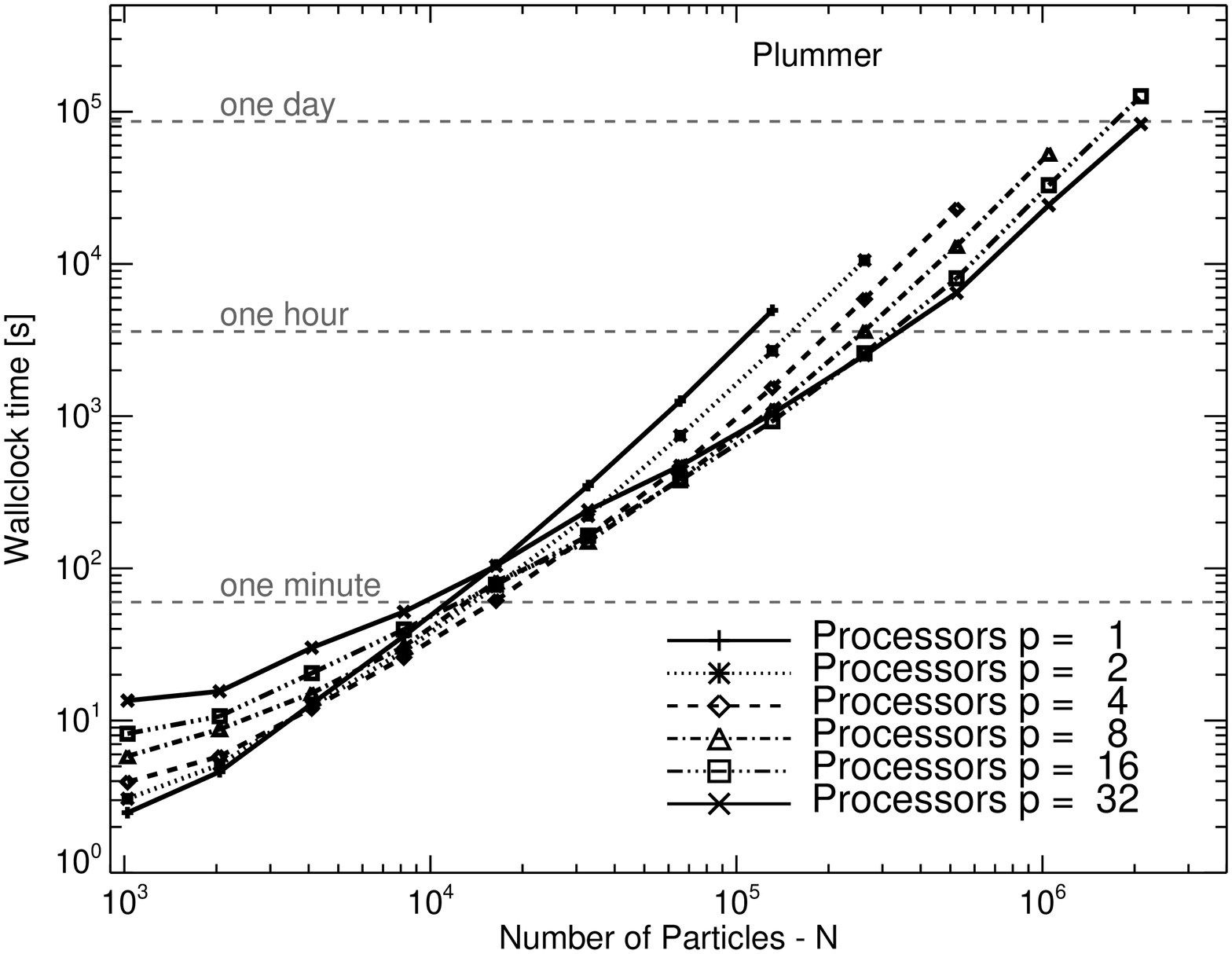} &
   \includegraphics[width=\hsize,angle=90]{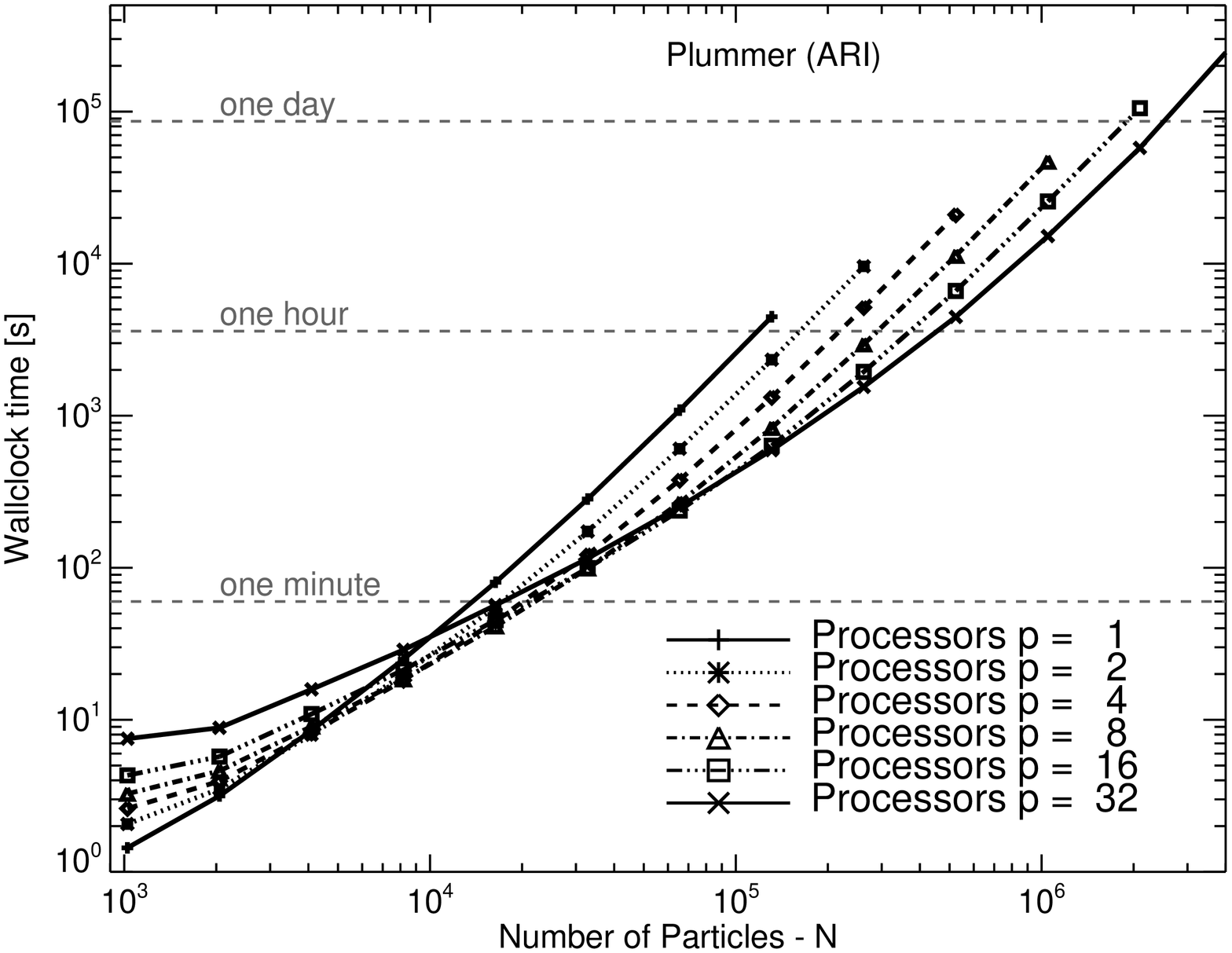} \\
   \includegraphics[width=\hsize,angle=90]{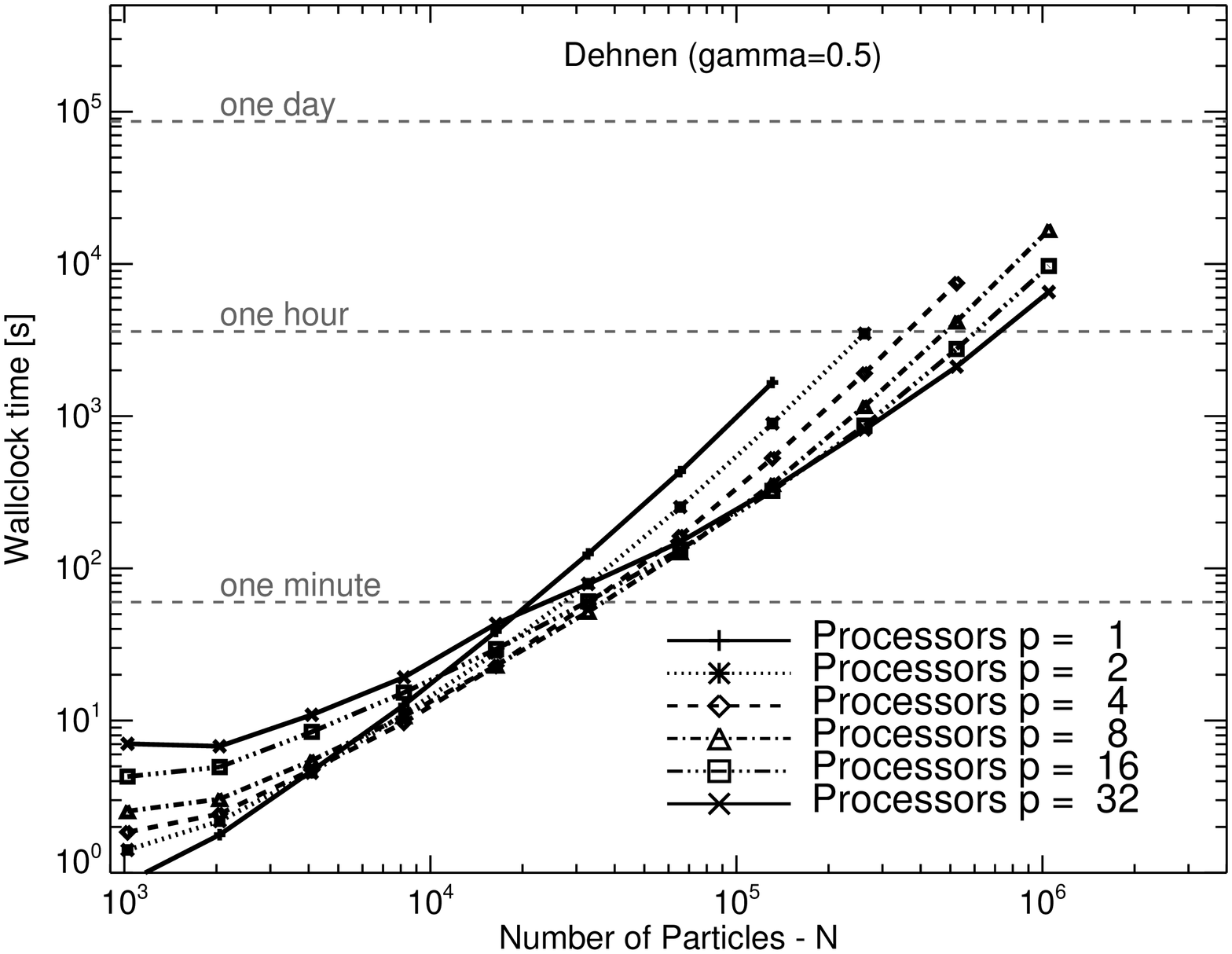} &
   \includegraphics[width=\hsize,angle=90]{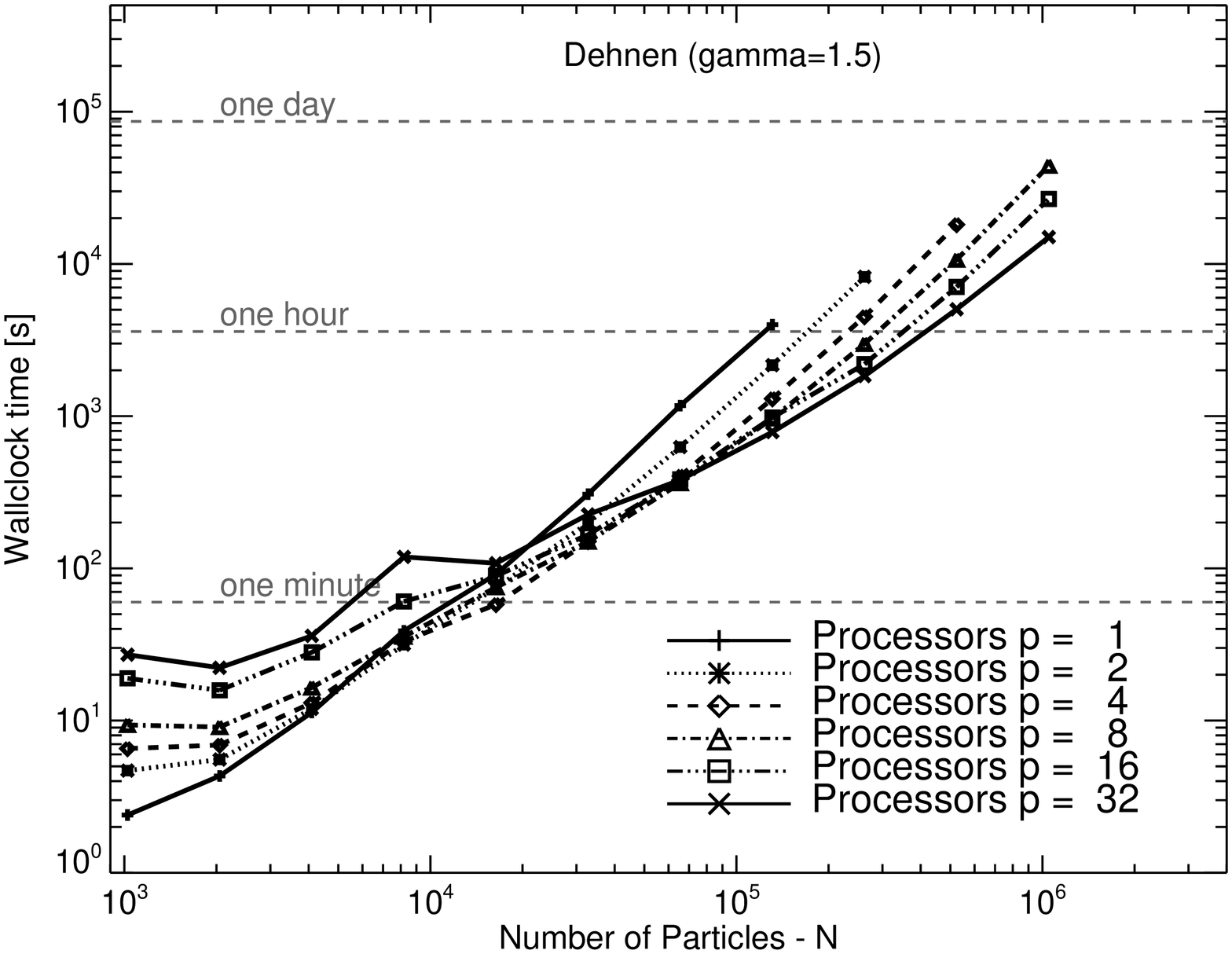} \\
   \includegraphics[width=\hsize,angle=90]{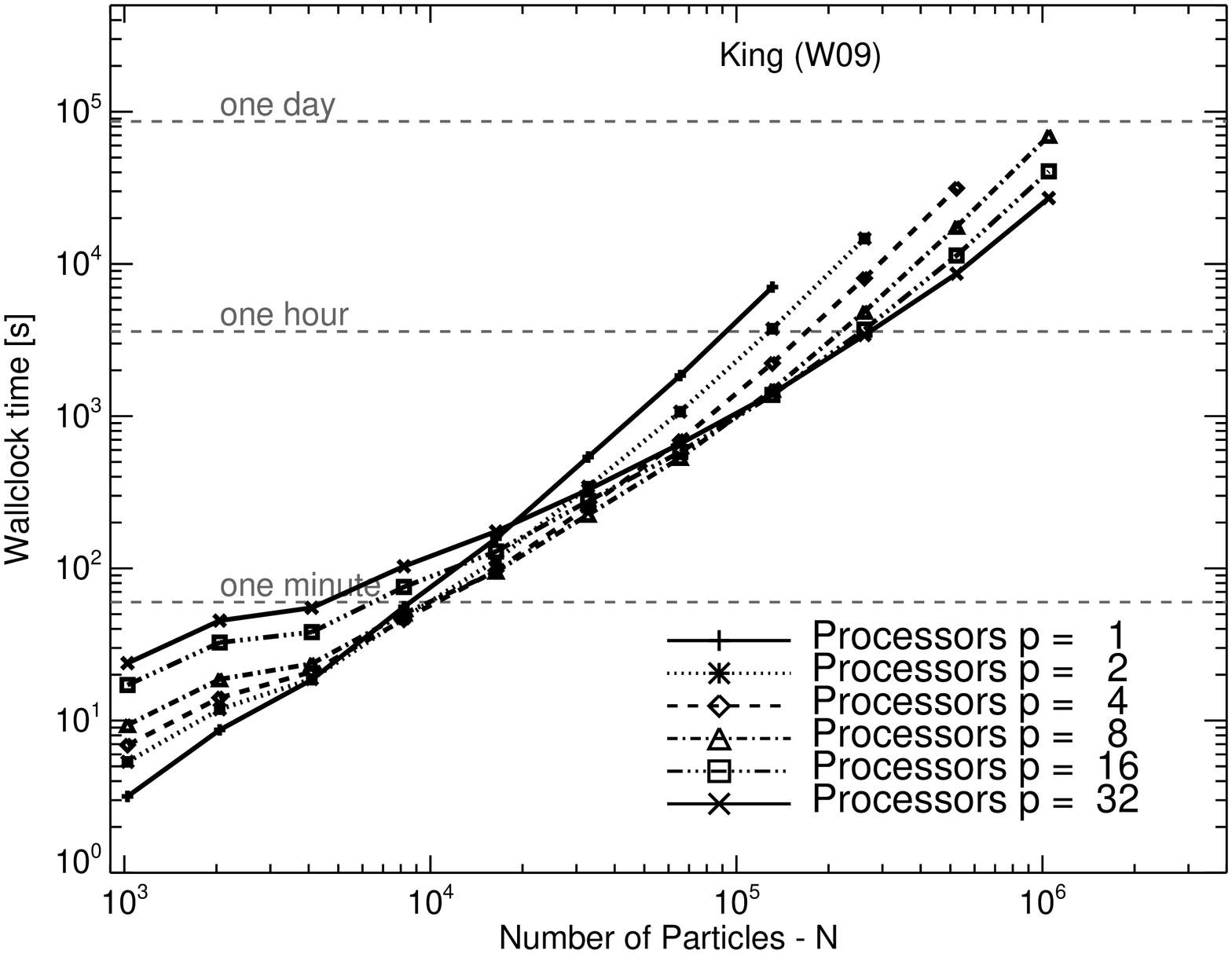} &
   \includegraphics[width=\hsize,angle=90]{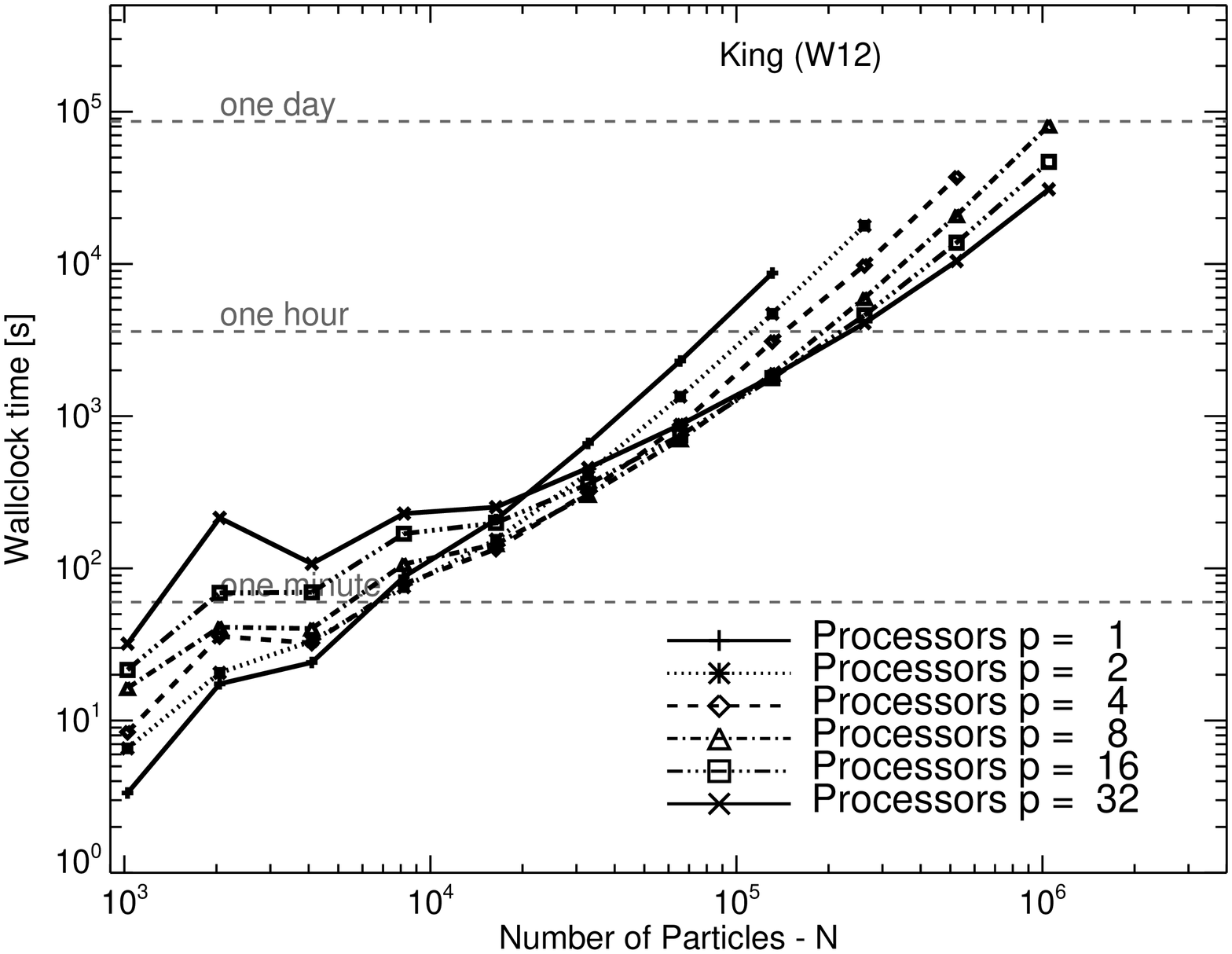} \\
   \end{tabular}} 
   \caption{Wallclock time $w$ versus particle number $N$ for different
            numbers of processors $p$. The plots in the top row show
            the results for a Plummer model on the RIT (left) and
            ARI (right) clusters. The remaining plots show wallclock
	    times of Dehnen models with $\gamma=0.5$ (middle left) and
            $\gamma=1.5$ (middle right) and of King models with
            $W_0=9$ (bottom left) and  $W_0=12$ (bottom right).}  
   \label{fig_ppwc} 
\end{figure}
%
%----------------------------------------------------------------------

%----------------------------------------------------------------------
%
%   Figure: Performance plots - Speedup
%   ===================================
%
\begin{figure*} 
   \resizebox{\hsize}{!}{  
   \begin{tabular}{cc}
   \includegraphics[width=\hsize,angle=90]{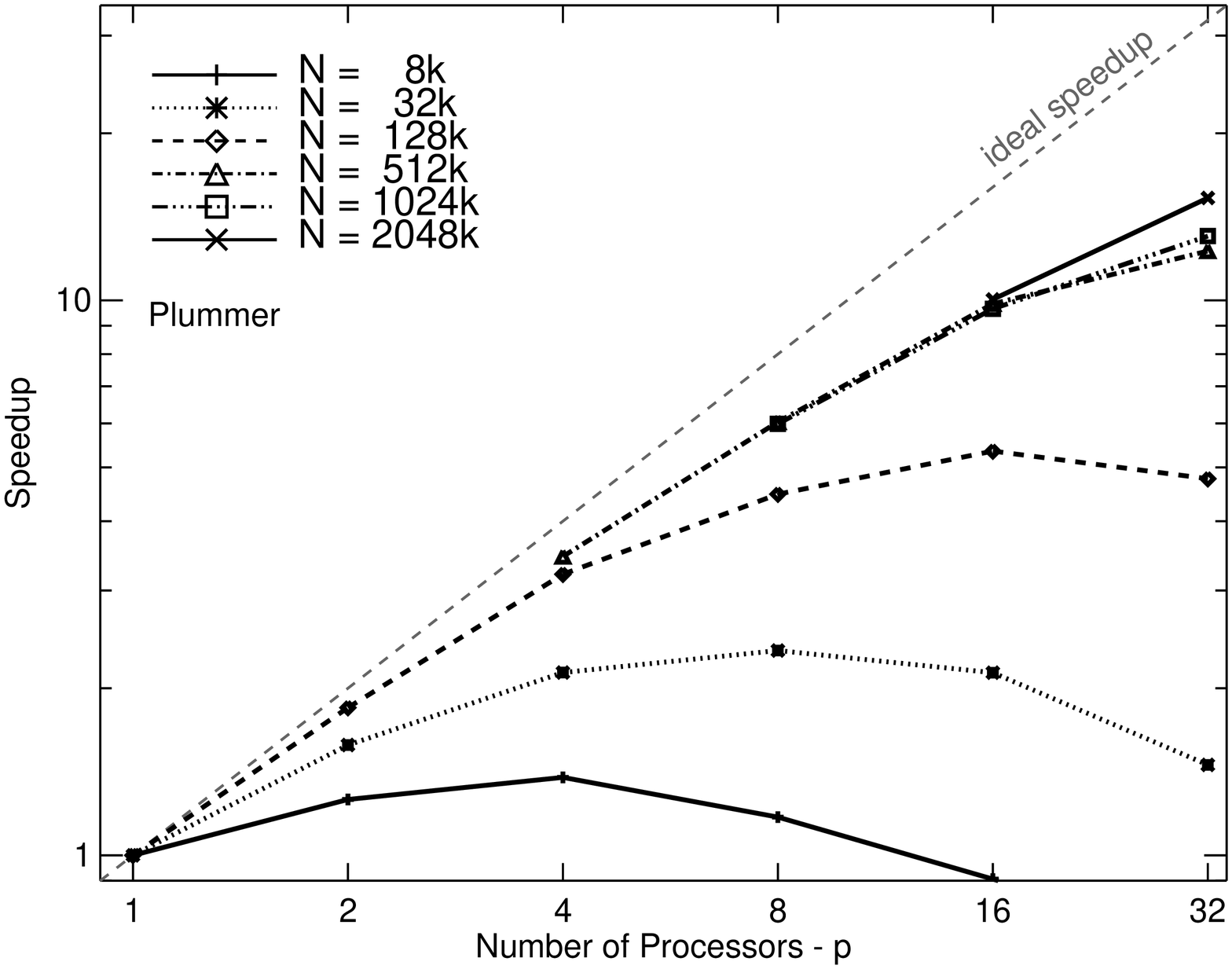} &
   \includegraphics[width=\hsize,angle=90]{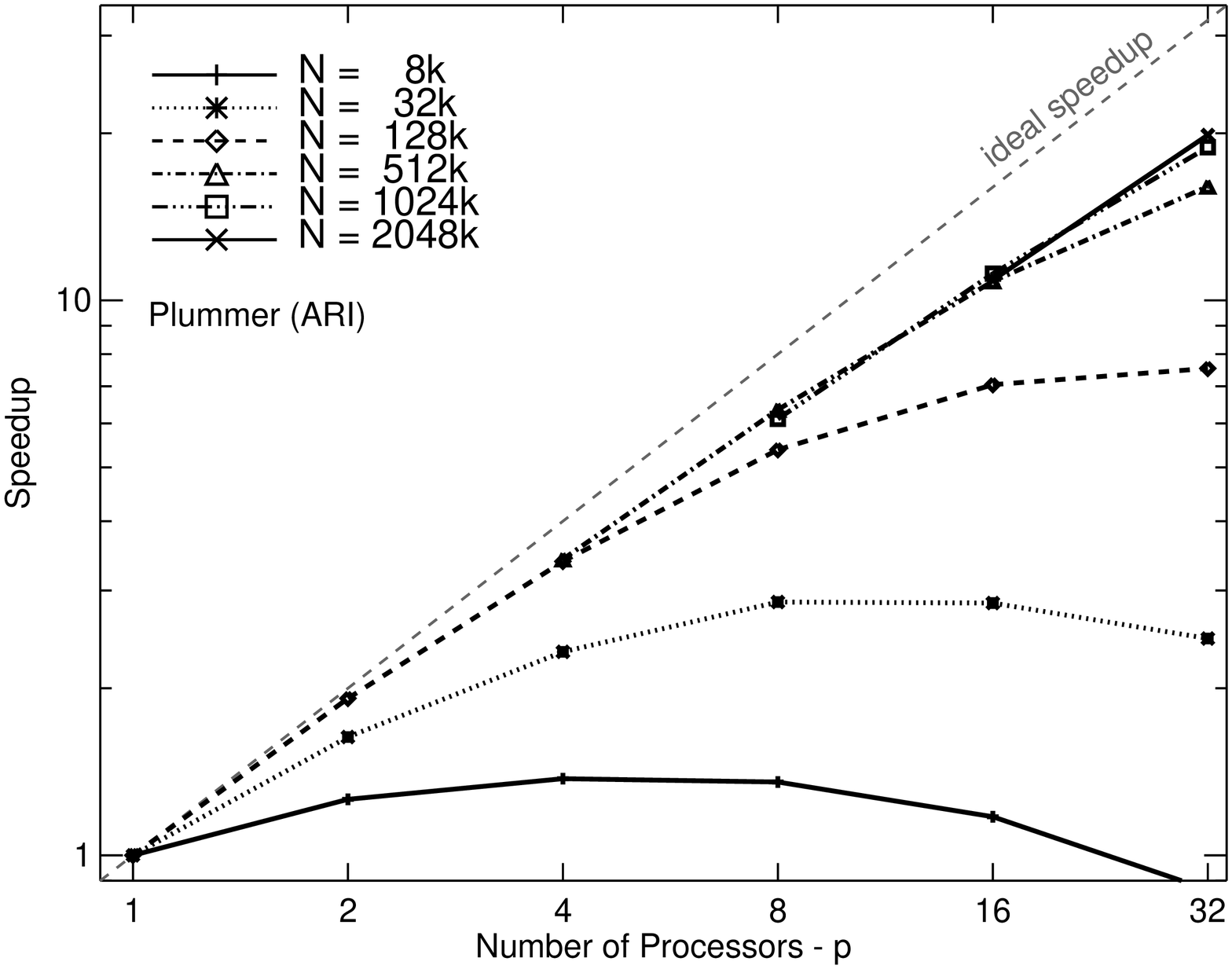} \\
   \includegraphics[width=\hsize,angle=90]{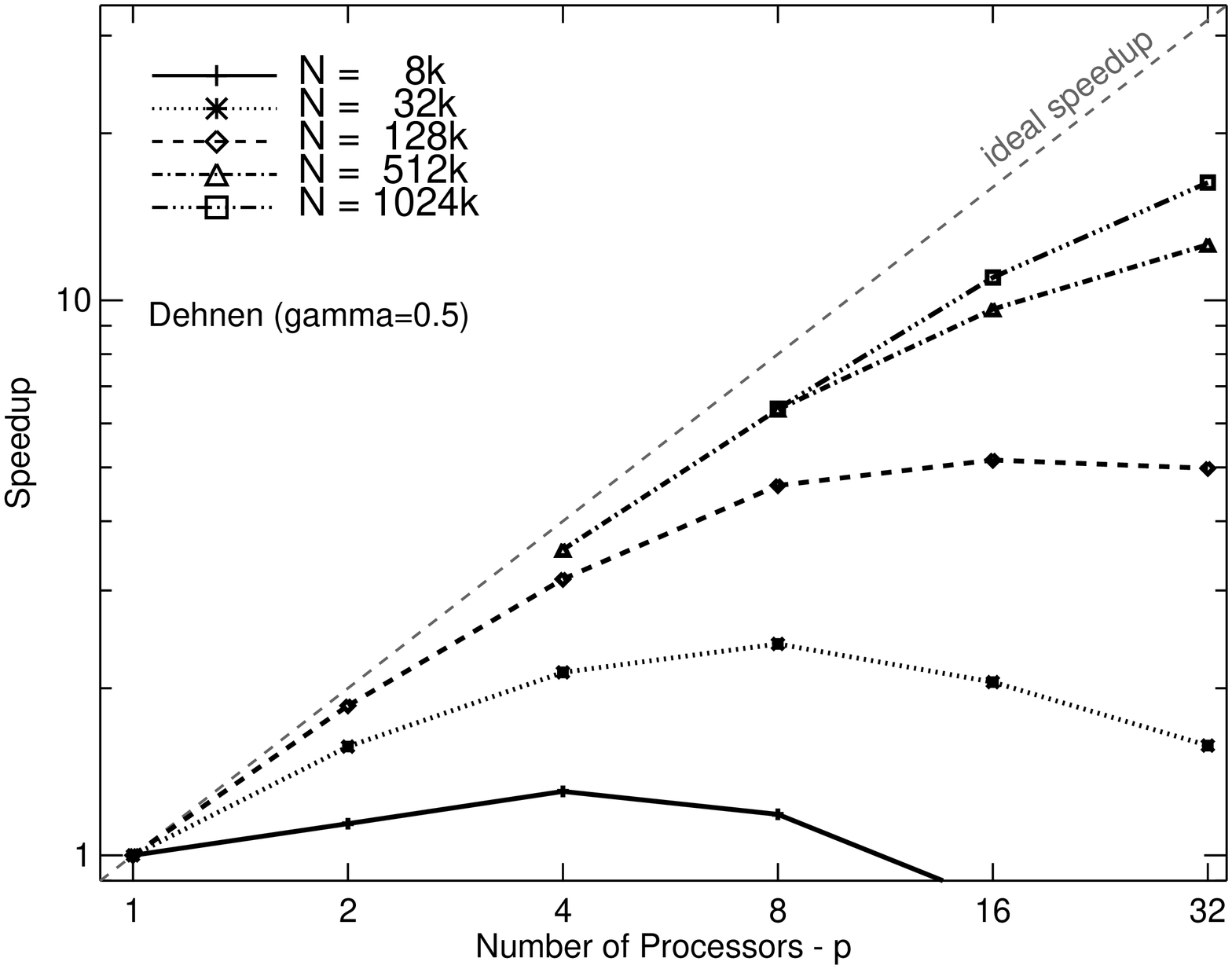} &
   \includegraphics[width=\hsize,angle=90]{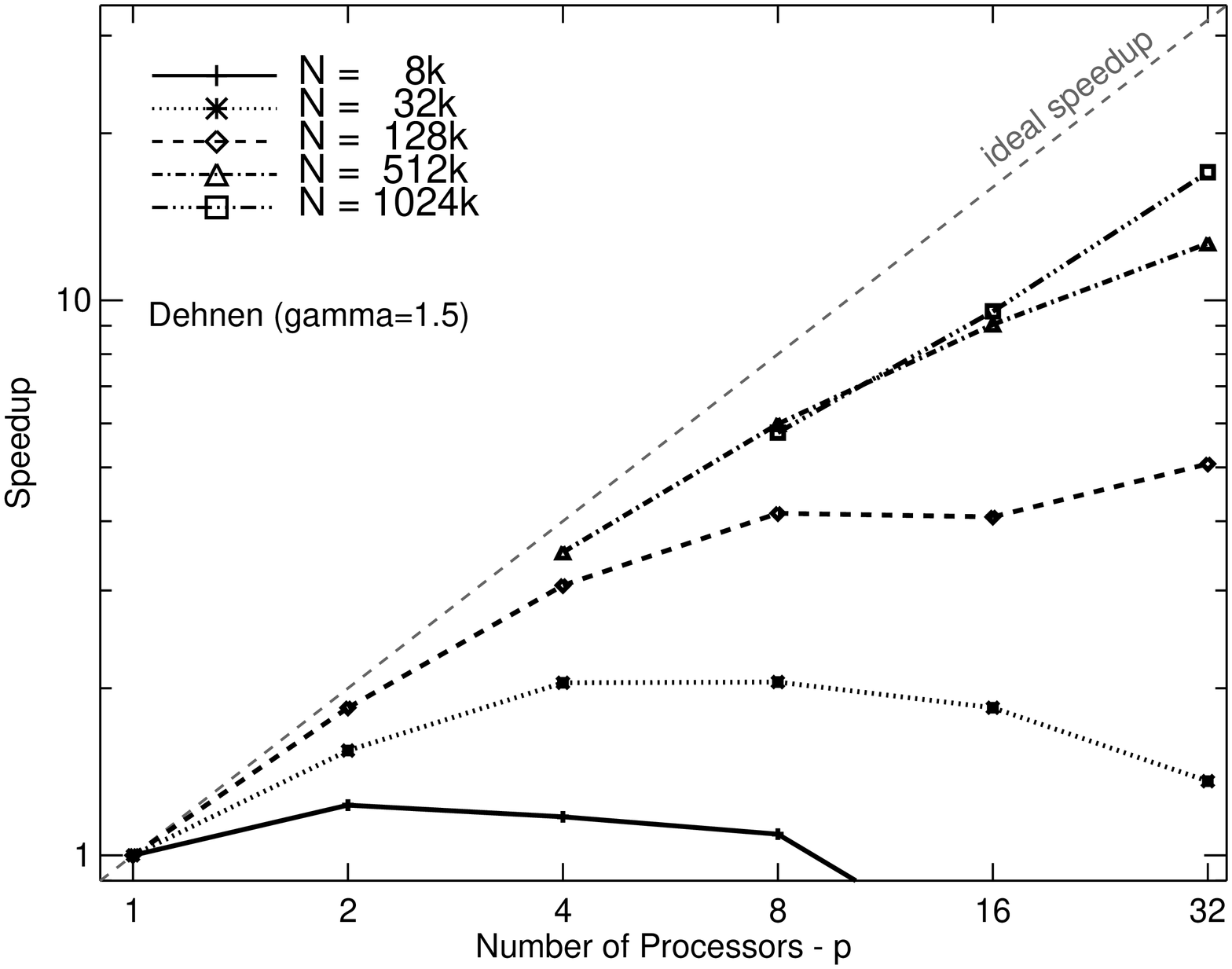} \\
   \includegraphics[width=\hsize,angle=90]{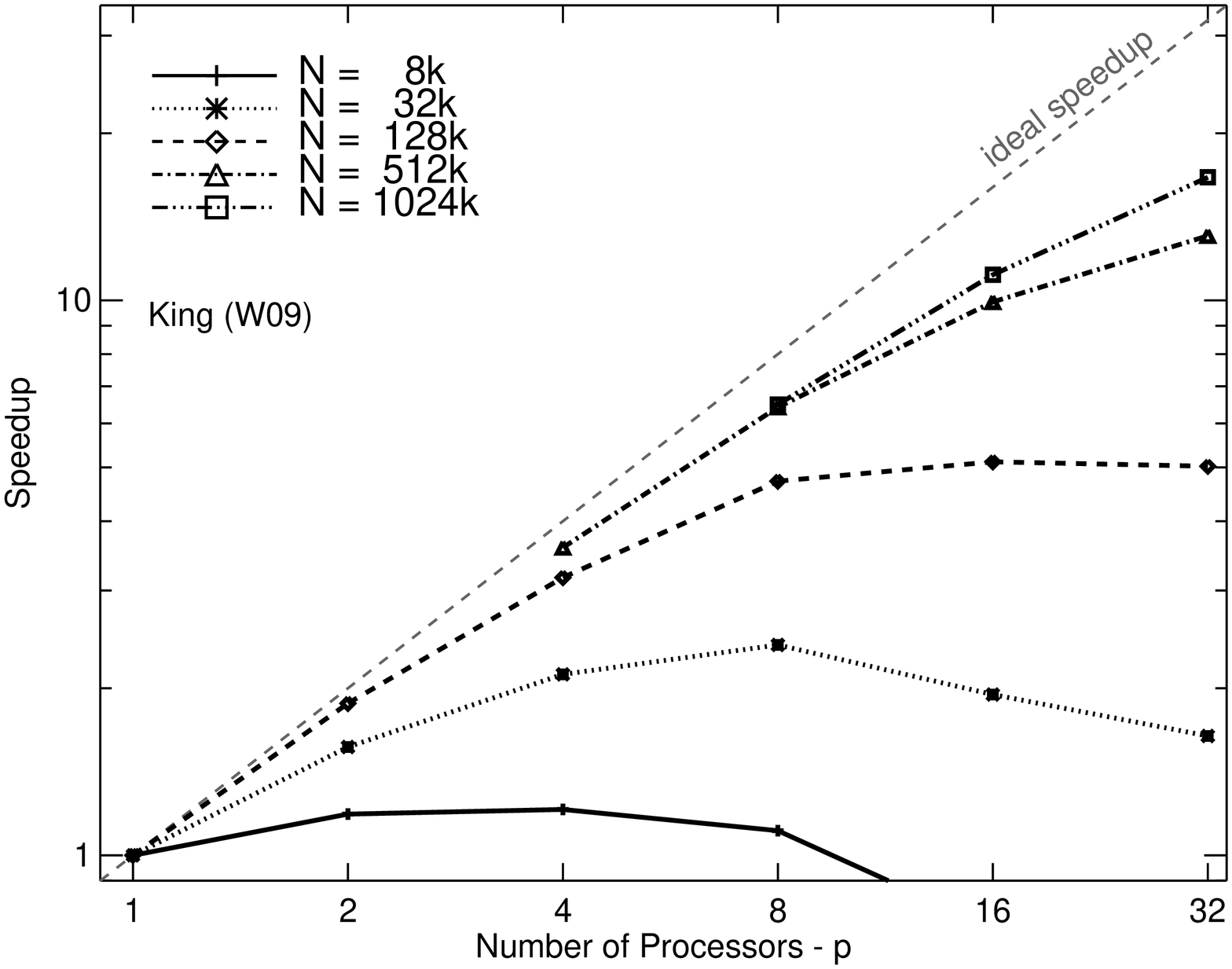} &
   \includegraphics[width=\hsize,angle=90]{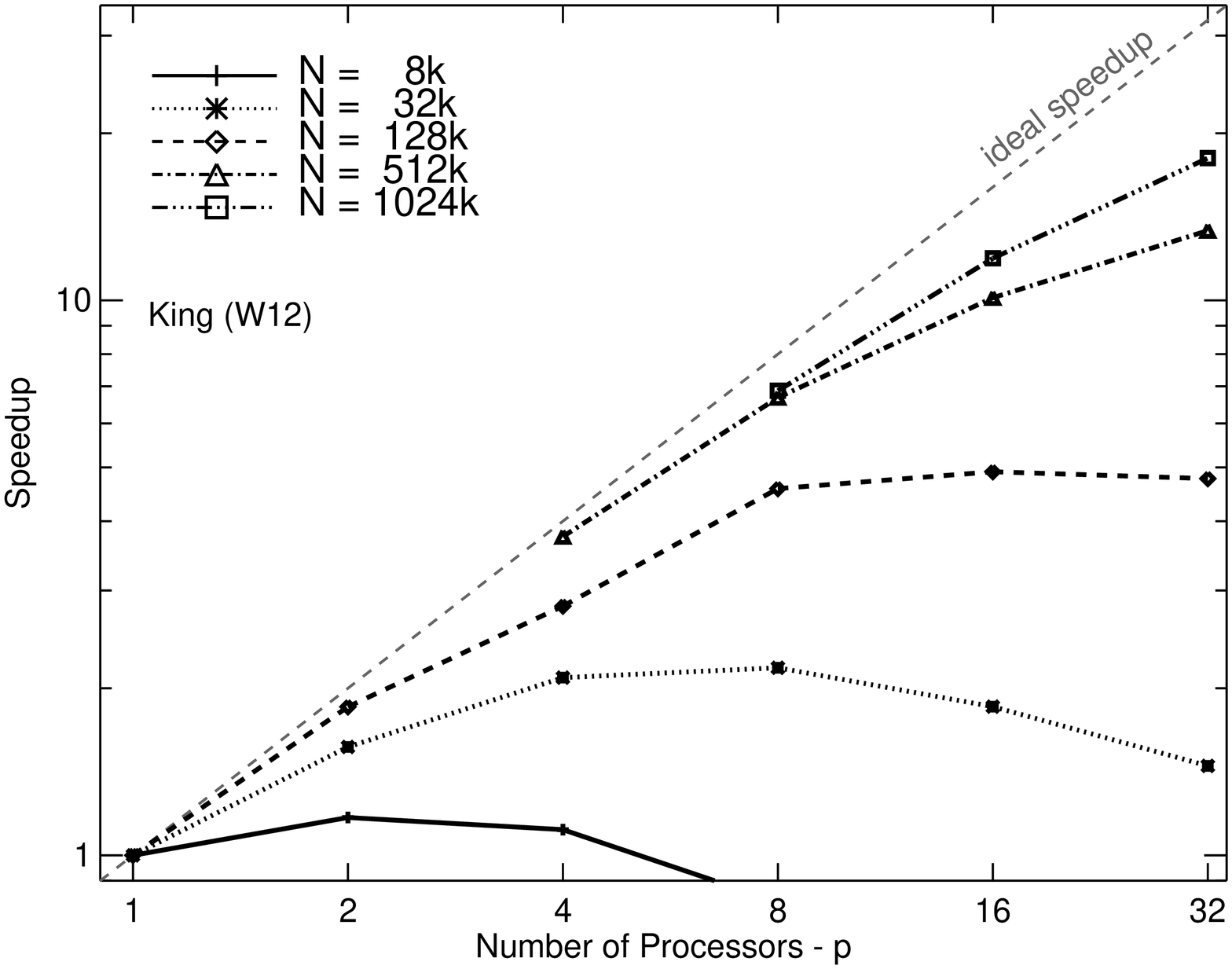} \\
   \end{tabular}} 
   \caption{Speedup $s$ versus processor number $p$ for different
            particle numbers $N$. The plots in the top row show
            the results for a Plummer model on the RIT (left) and 
            ARI (right) clusters. The remaining plots show the speedup
	    for Dehnen models with $\gamma=0.5$ (middle left) and
            $\gamma=1.5$ (middle right) and for King models with
            $W_0=9$ (bottom left) and  $W_0=12$ (bottom right).}  
   \label{fig_ppsu} 
\end{figure*}
%
%----------------------------------------------------------------------

%----------------------------------------------------------------------
%
%   Figure: Performance plots - Efficiency
%   ======================================
%
\begin{figure*} 
   \resizebox{\hsize}{!}{  
   \begin{tabular}{cc}
   \includegraphics[width=\hsize,angle=90]{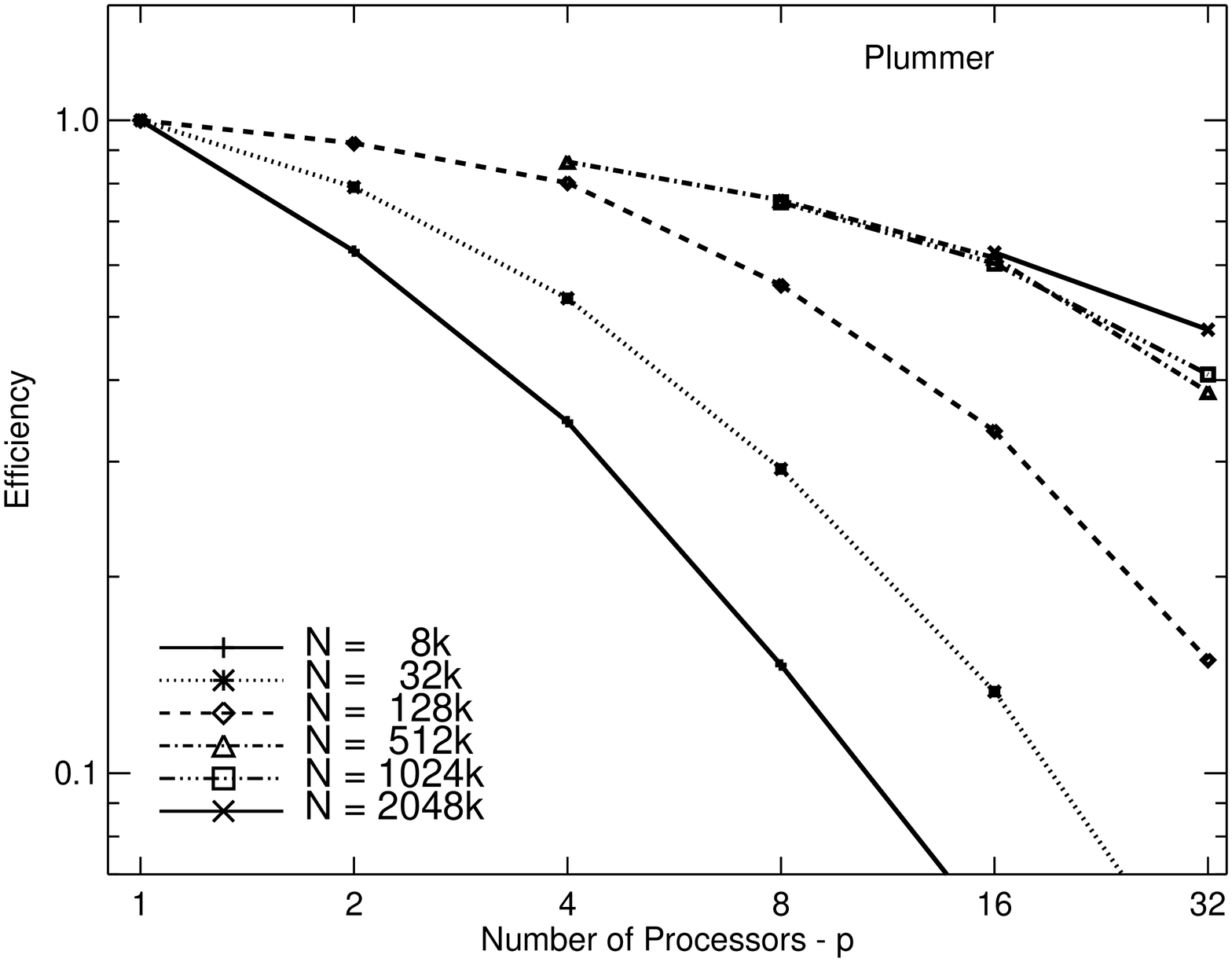} &
   \includegraphics[width=\hsize,angle=90]{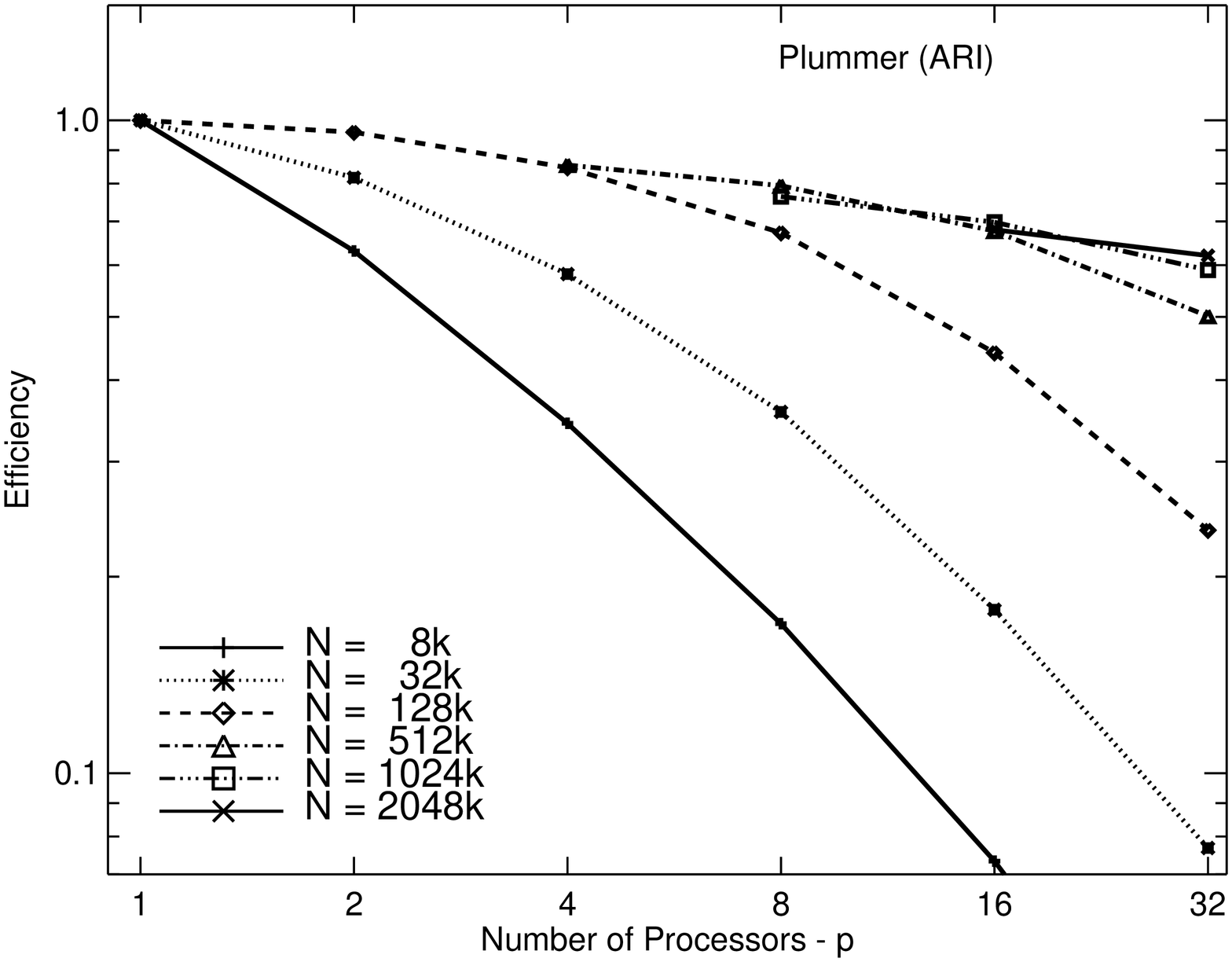} \\
   \includegraphics[width=\hsize,angle=90]{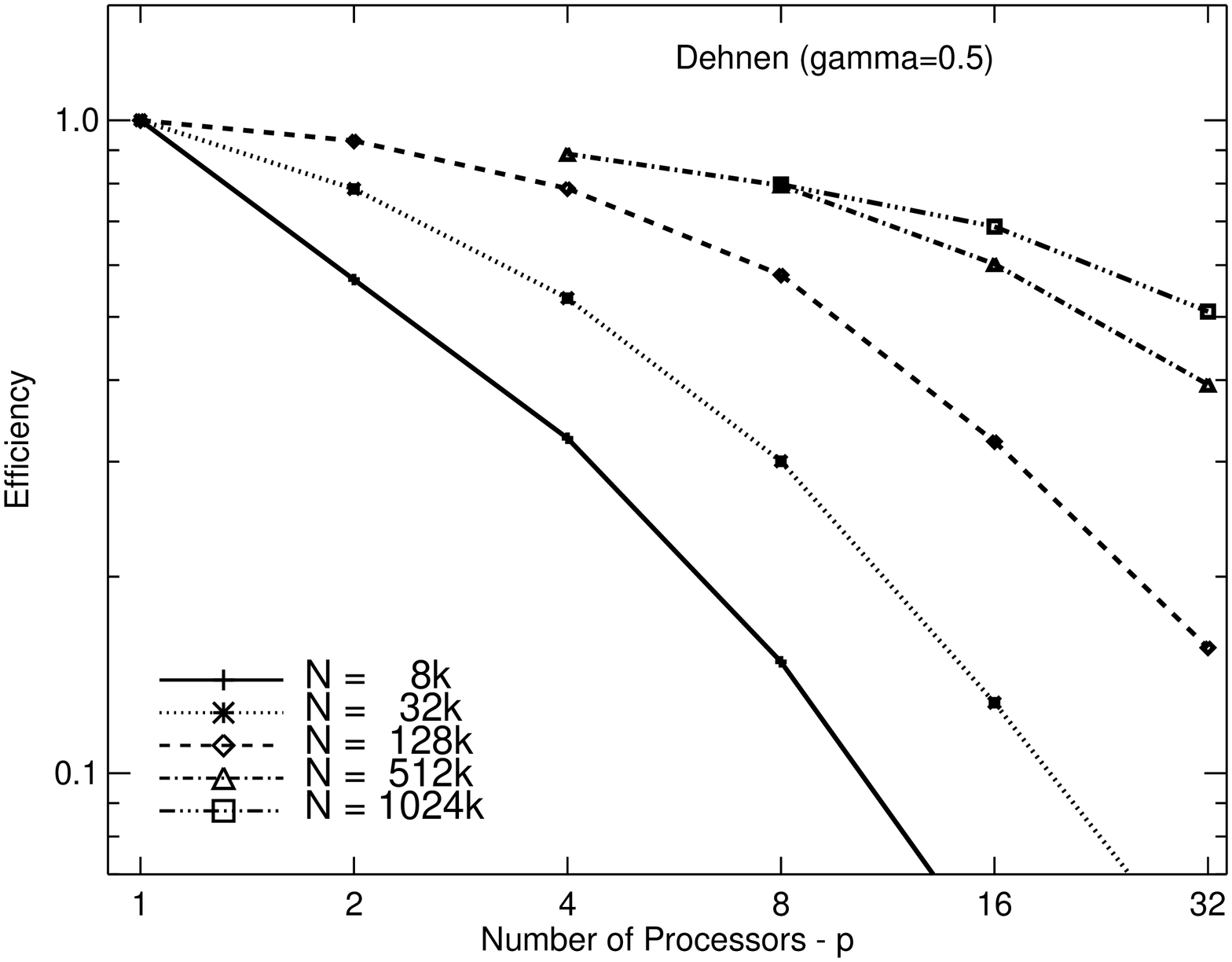} &
   \includegraphics[width=\hsize,angle=90]{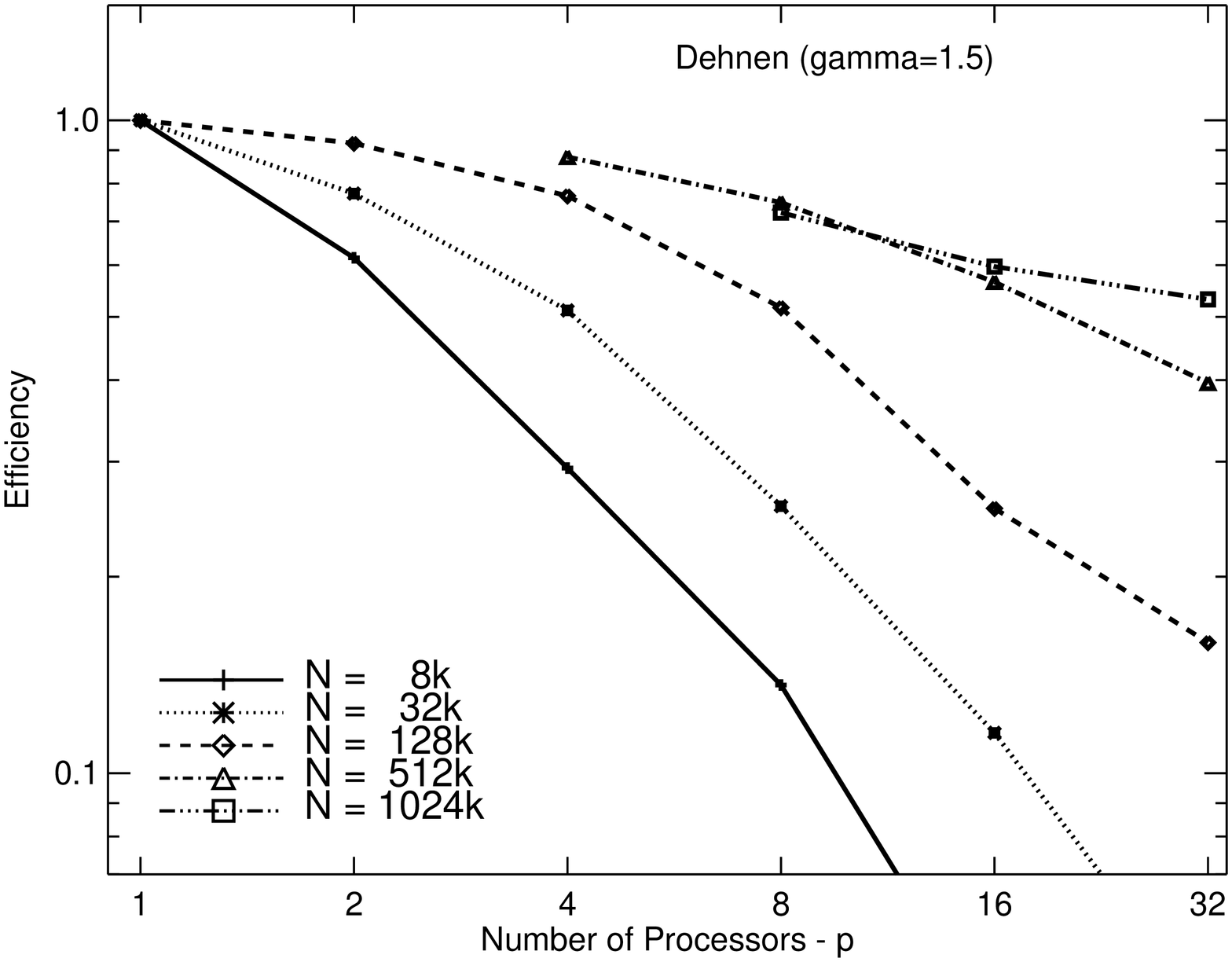} \\
   \includegraphics[width=\hsize,angle=90]{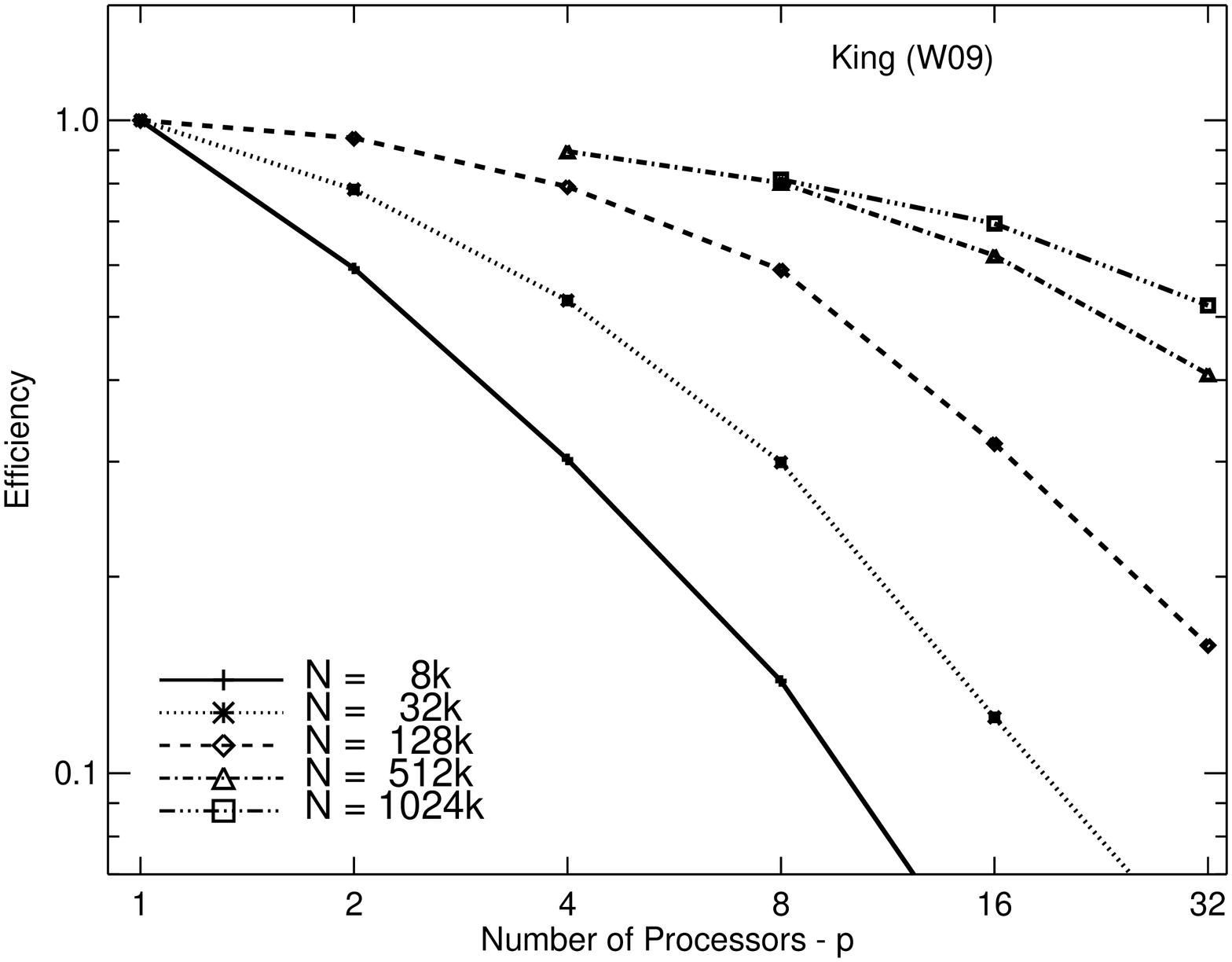} &
   \includegraphics[width=\hsize,angle=90]{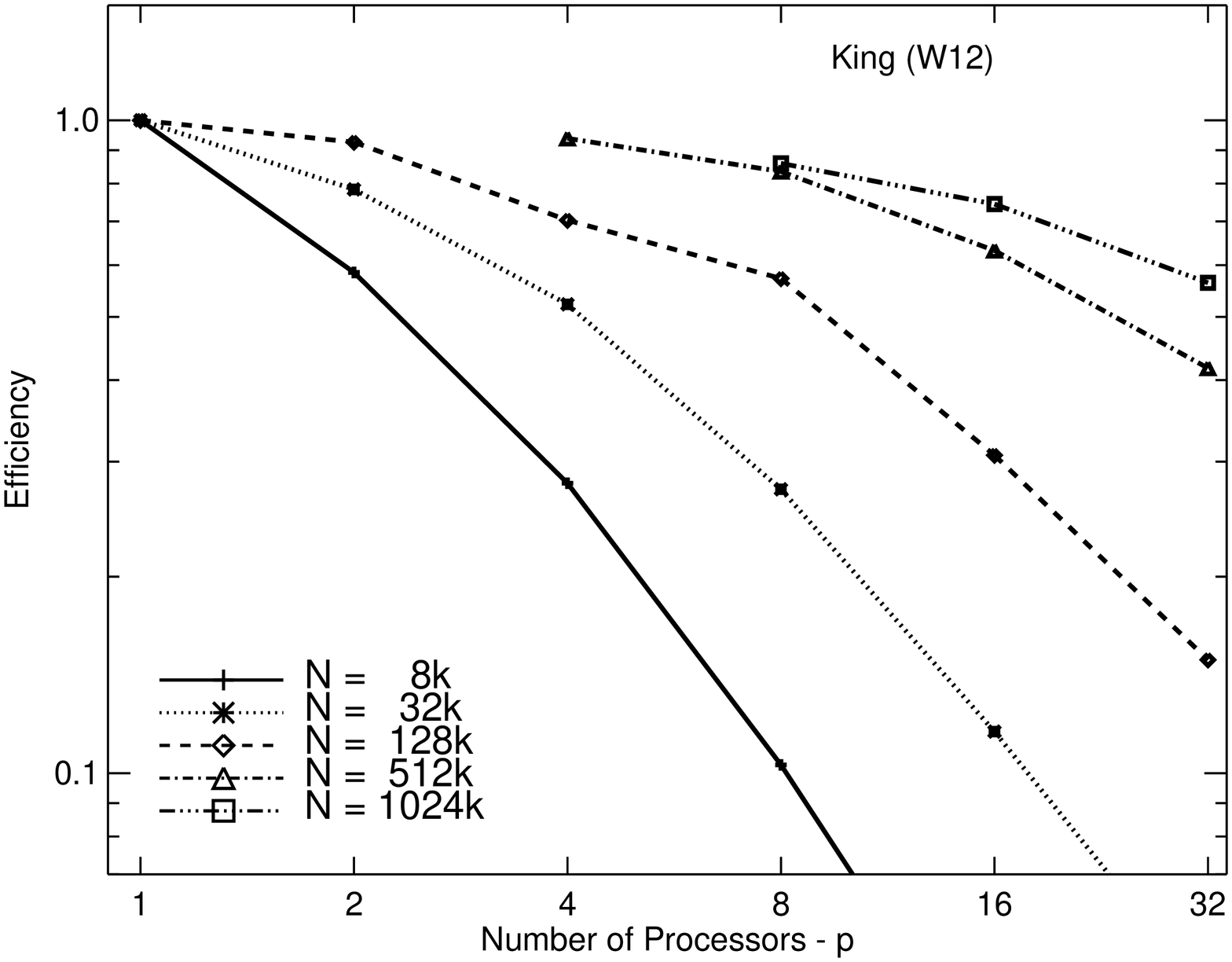} \\
   \end{tabular}} 
   \caption{Efficiency $e$ vs. number of processors $p$ for different
            particle numbers $N$. The plots in the top row show
            the results for a Plummer model on the RIT (left) and
            ARI (right) clusters. The remaining plots show efficiencies
	    for Dehnen models with $\gamma=0.5$ (middle left) and
            $\gamma=1.5$ (middle right) and for King models with
            $W_0=9$ (bottom left) and  $W_0=12$ (bottom right).}  
   \label{fig_ppeff} 
\end{figure*}
%
%----------------------------------------------------------------------

%----------------------------------------------------------------------
%
%   Figure: Performance plots - Speed
%   =================================
%
\begin{figure*} 
   \resizebox{\hsize}{!}{  
   \begin{tabular}{cc}
   \includegraphics[width=\hsize,angle=90]{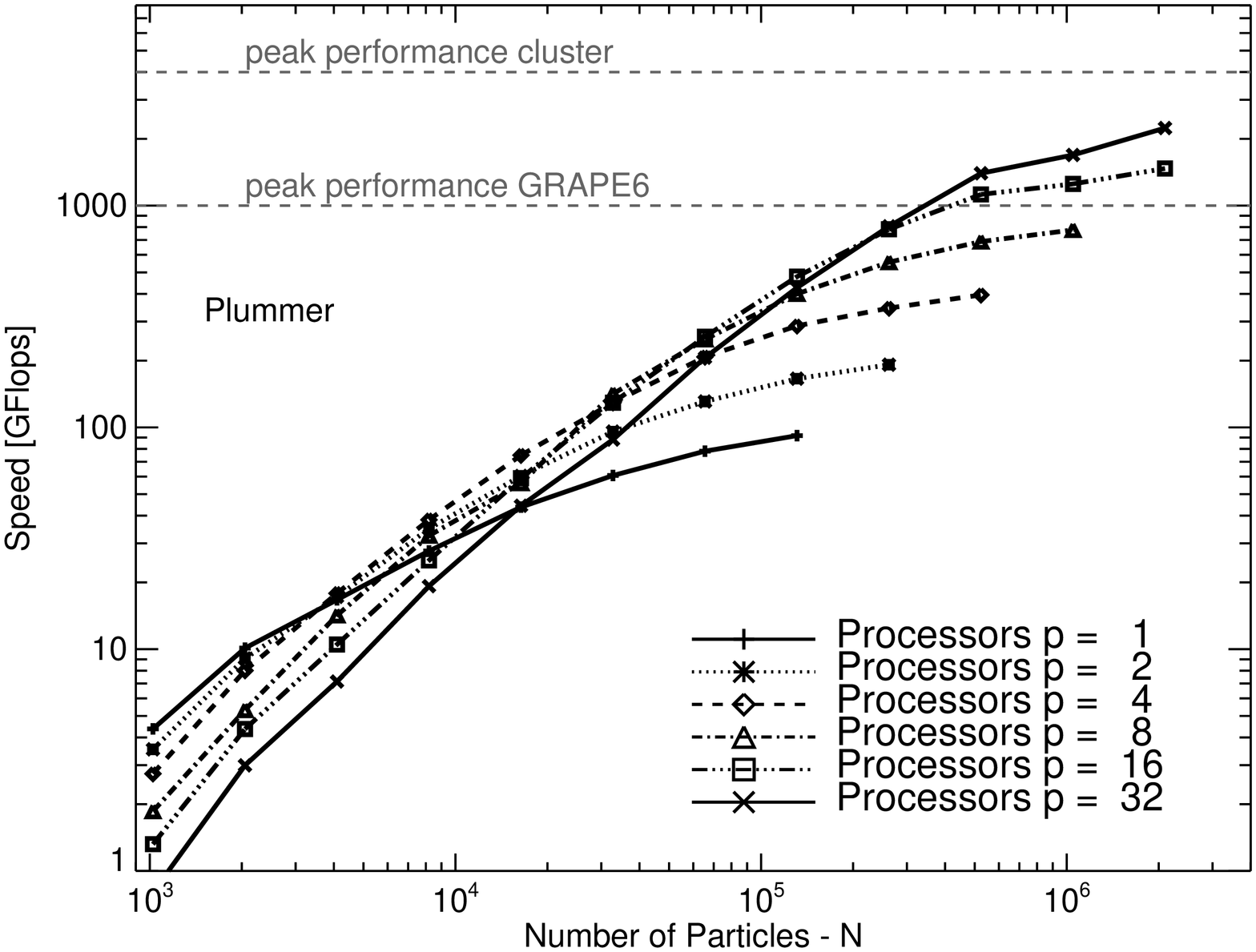} &
   \includegraphics[width=\hsize,angle=90]{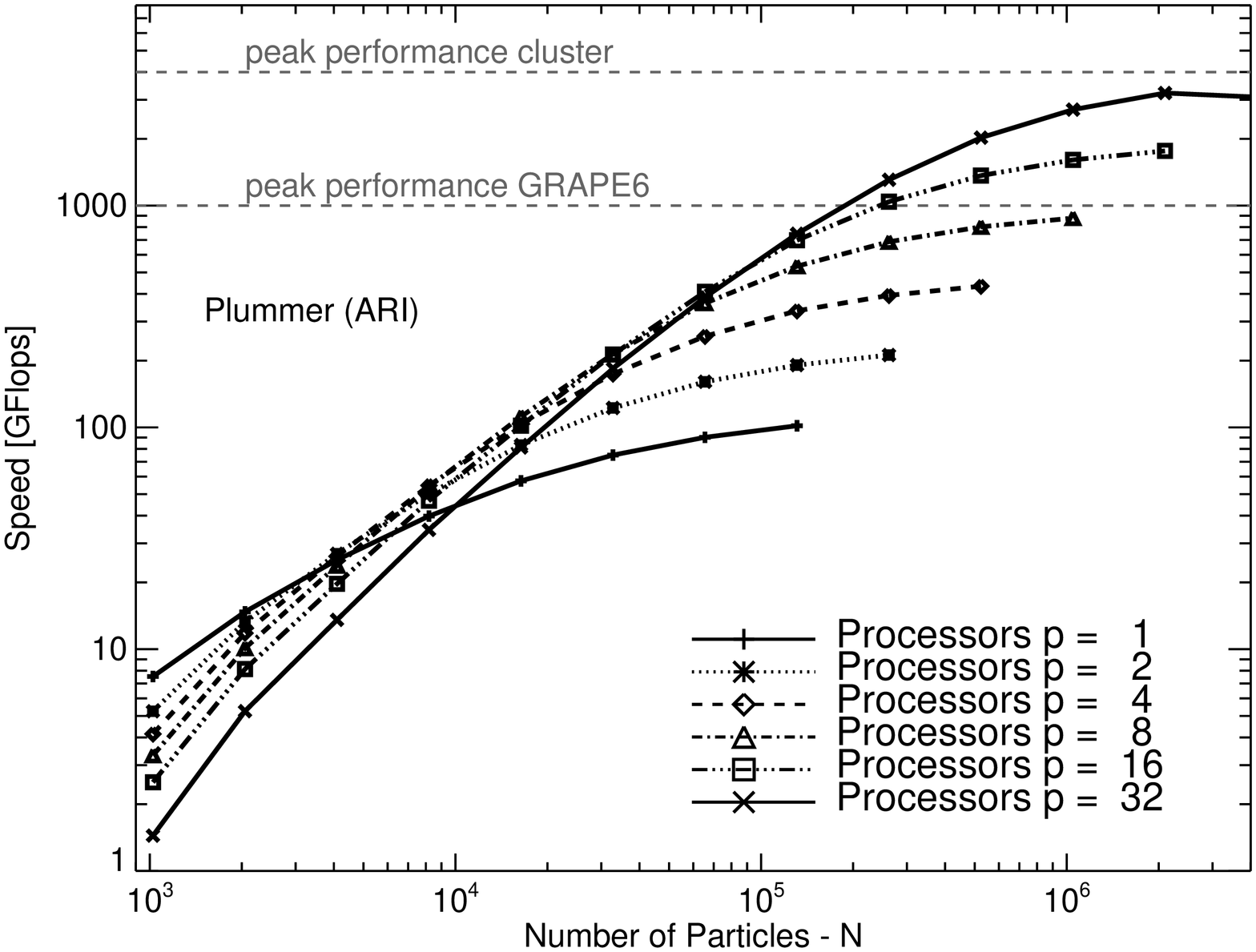} \\
   \includegraphics[width=\hsize,angle=90]{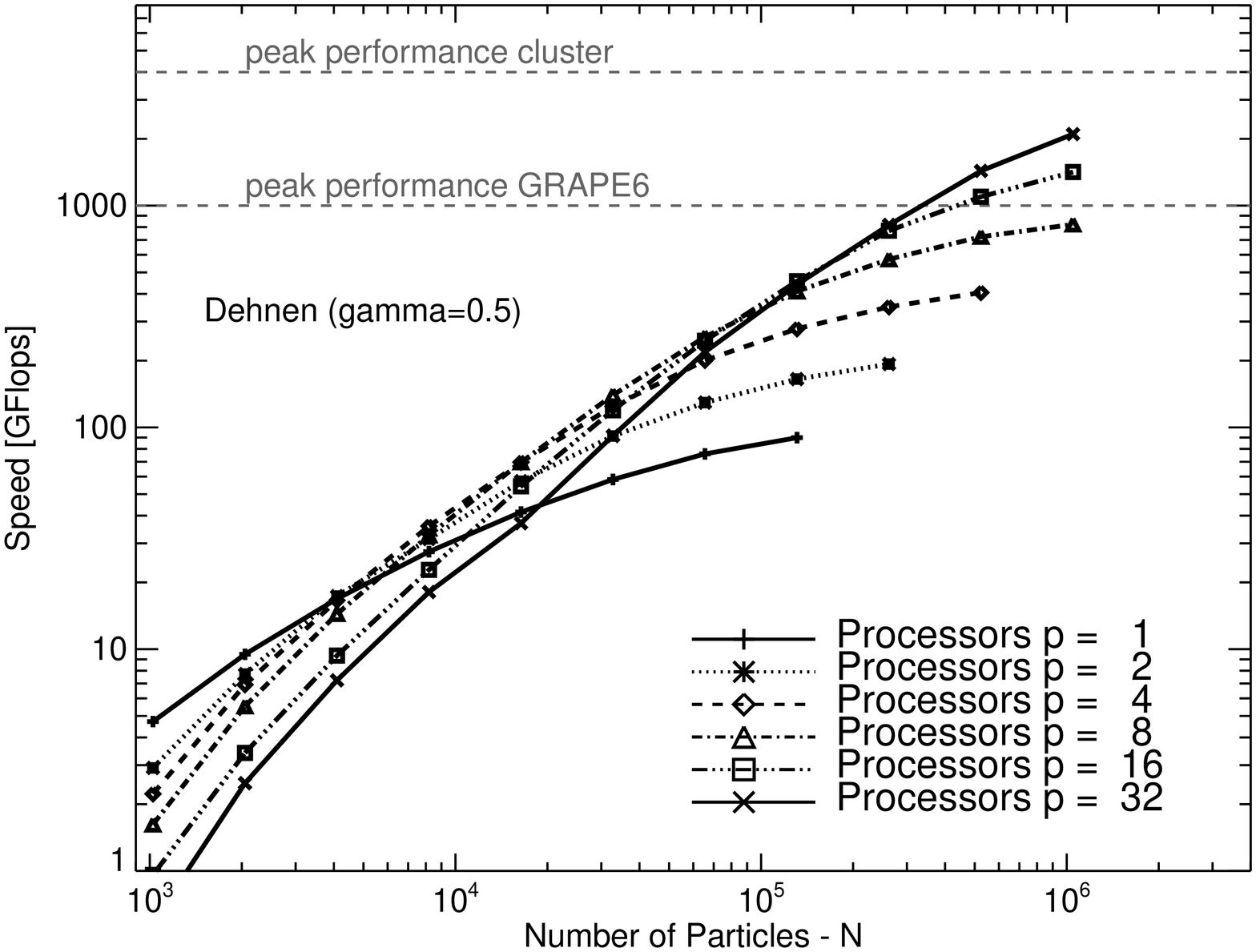} &
   \includegraphics[width=\hsize,angle=90]{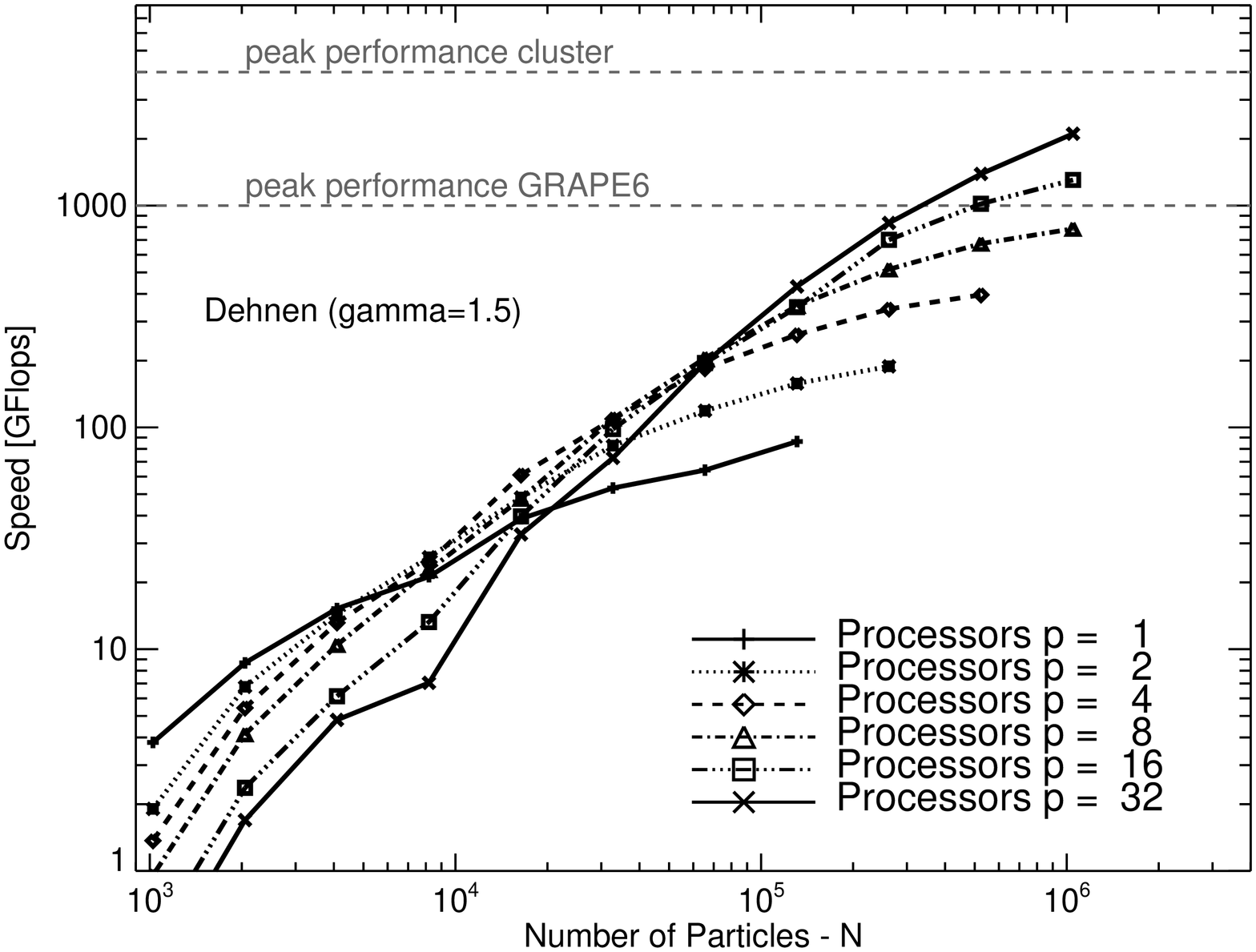} \\
   \includegraphics[width=\hsize,angle=90]{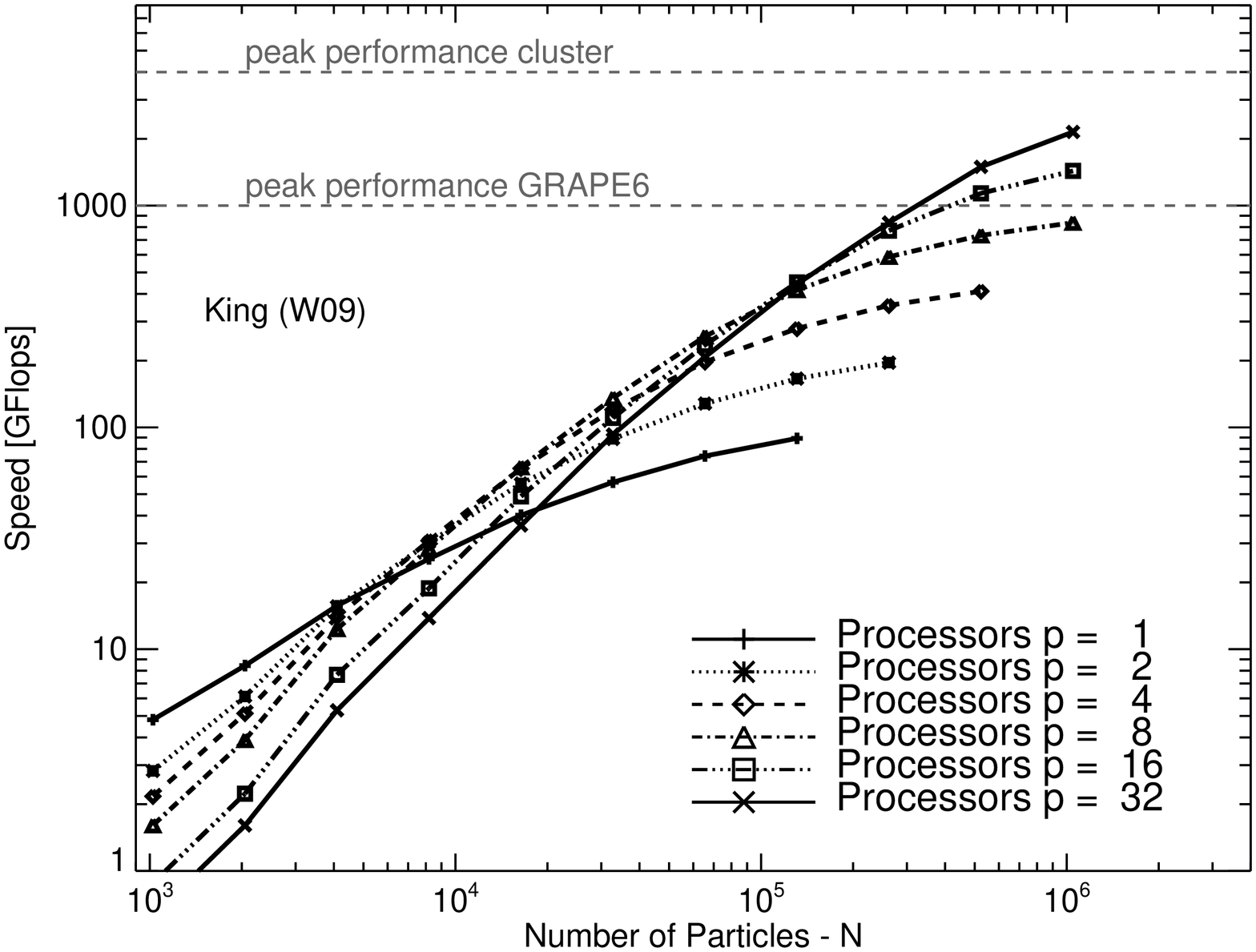} &
   \includegraphics[width=\hsize,angle=90]{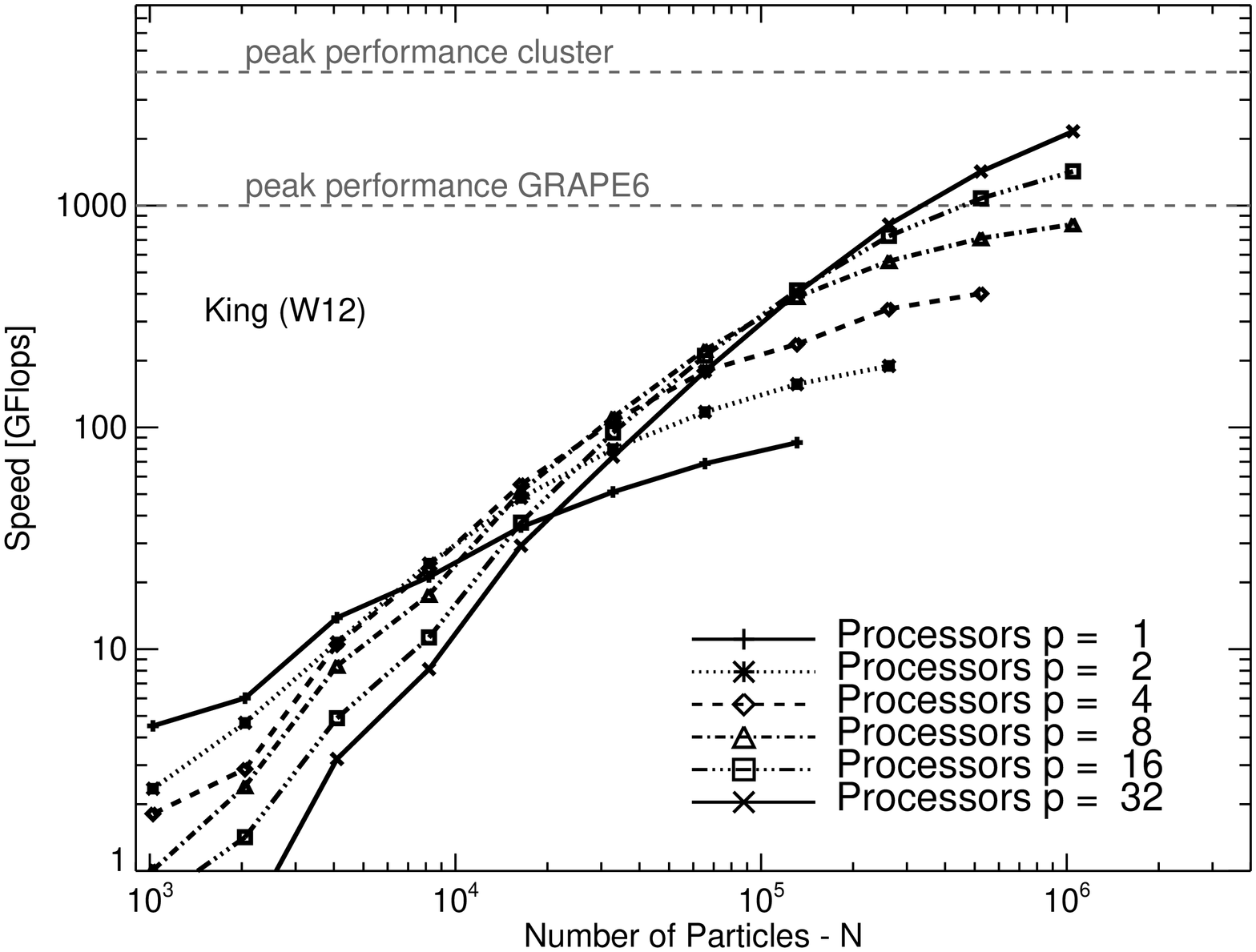} \\
   \end{tabular}} 
   \caption{Speed $f$ vs. particle number $N$ for different
            numbers of processors $p$. The plots in the top row show
            the results for a Plummer model on the RIT (left) and 
            ARI (right) clusters. The remaining plots show the speed
	    for Dehnen models with $\gamma=0.5$ (middle left) and
            $\gamma=1.5$ (middle right) and for King models with
            $W_0=9$ (bottom left) and  $W_0=12$ (bottom right).}  
   \label{fig_ppsp} 
\end{figure*}
%
%----------------------------------------------------------------------

%----------------------------------------------------------------------
%
%   Figure: Performance plots - Speed ratio
%   =======================================
%
\begin{figure*} 
   \resizebox{\hsize}{!}{  
   \begin{tabular}{cc}
   \includegraphics[width=\hsize,angle=90]{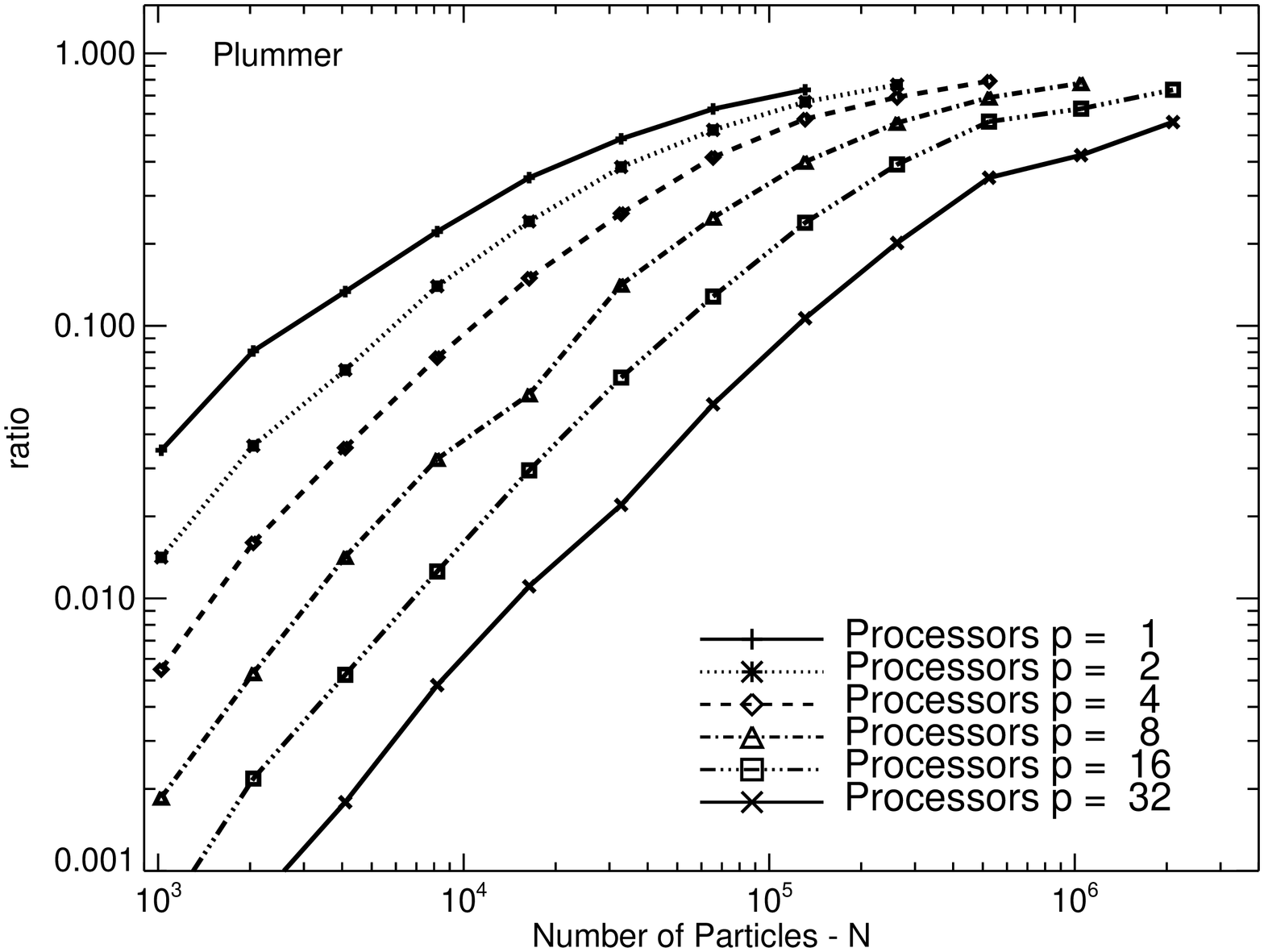} &
   \includegraphics[width=\hsize,angle=90]{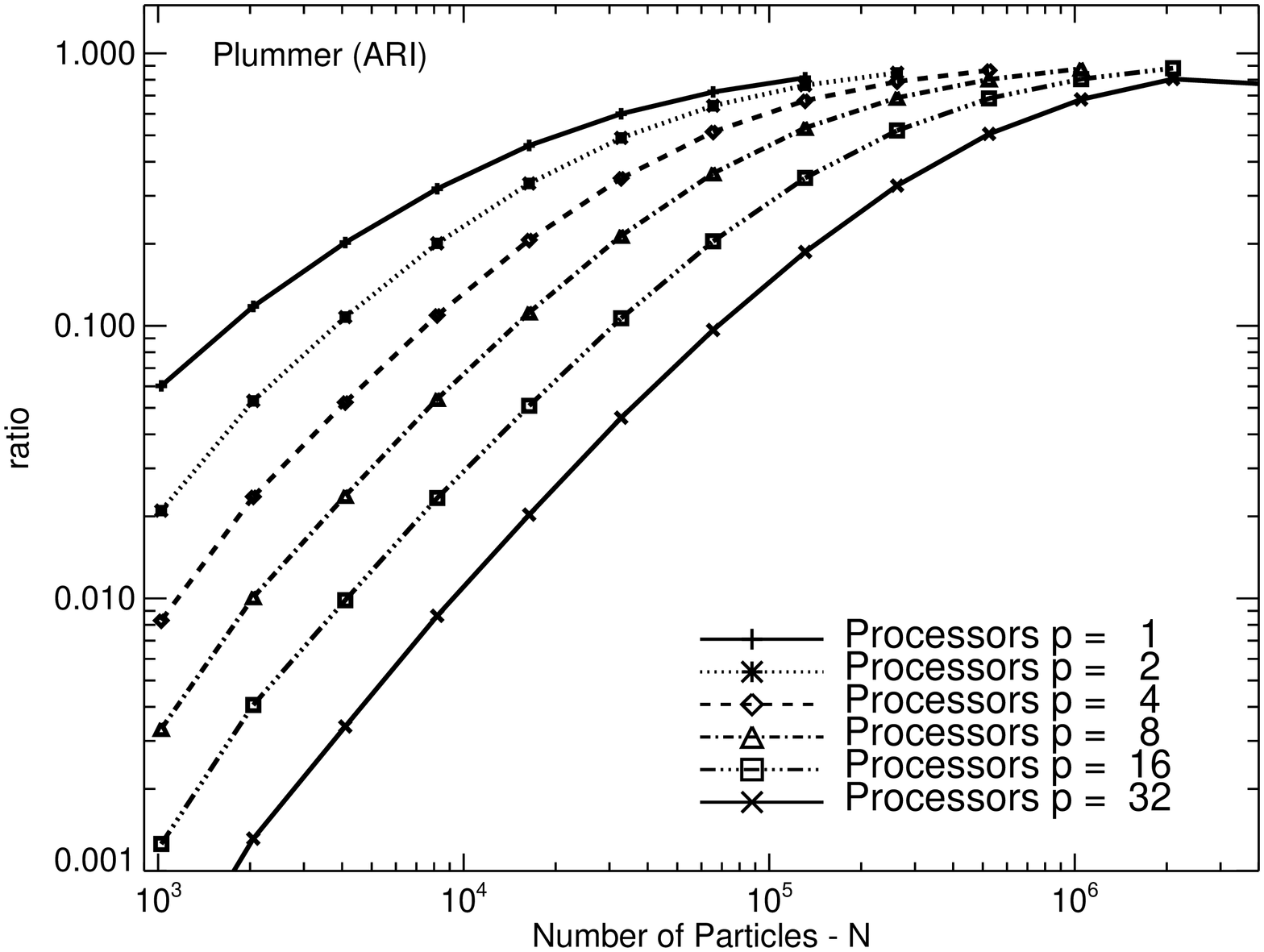} \\
   \includegraphics[width=\hsize,angle=90]{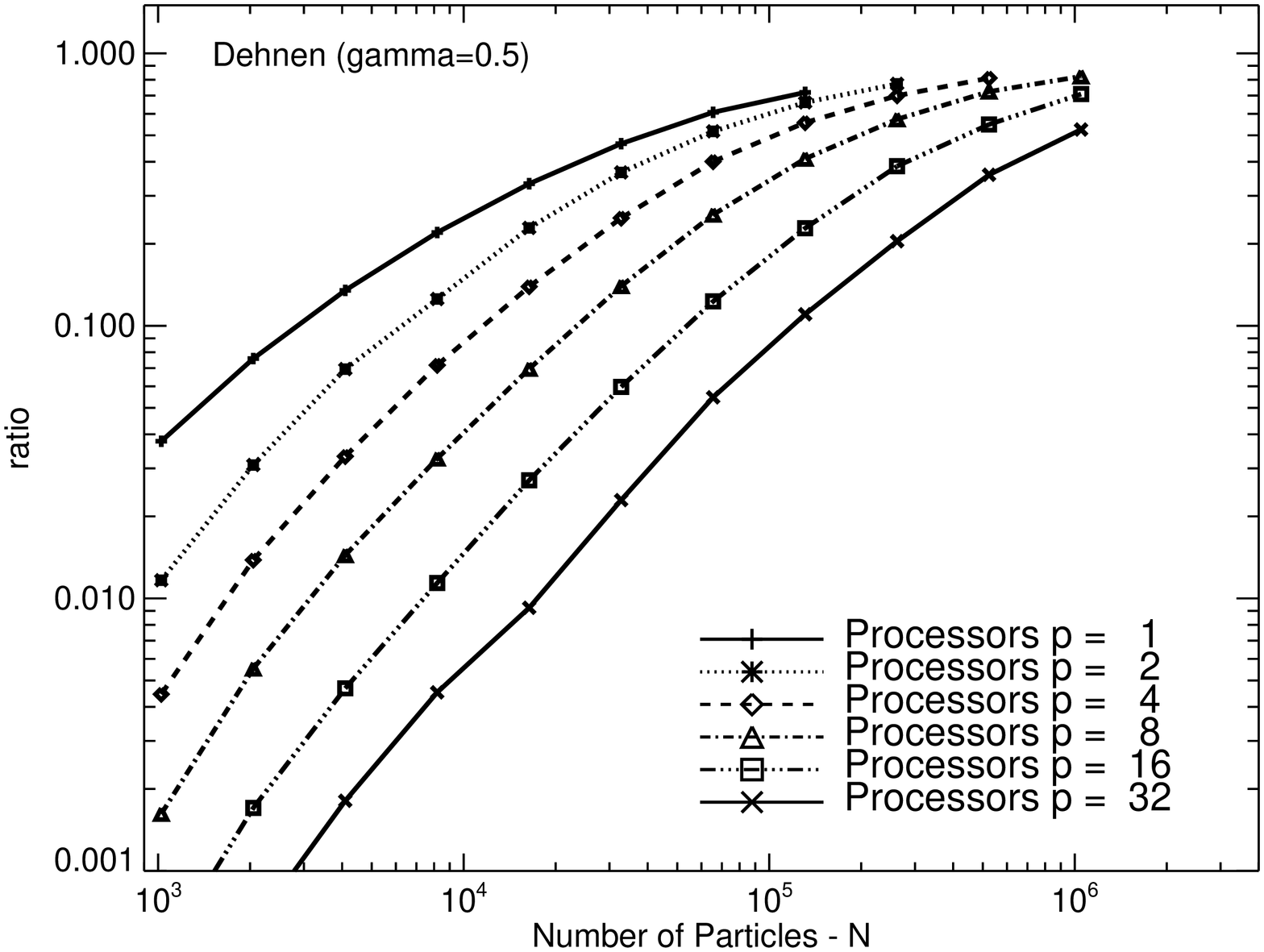} &
   \includegraphics[width=\hsize,angle=90]{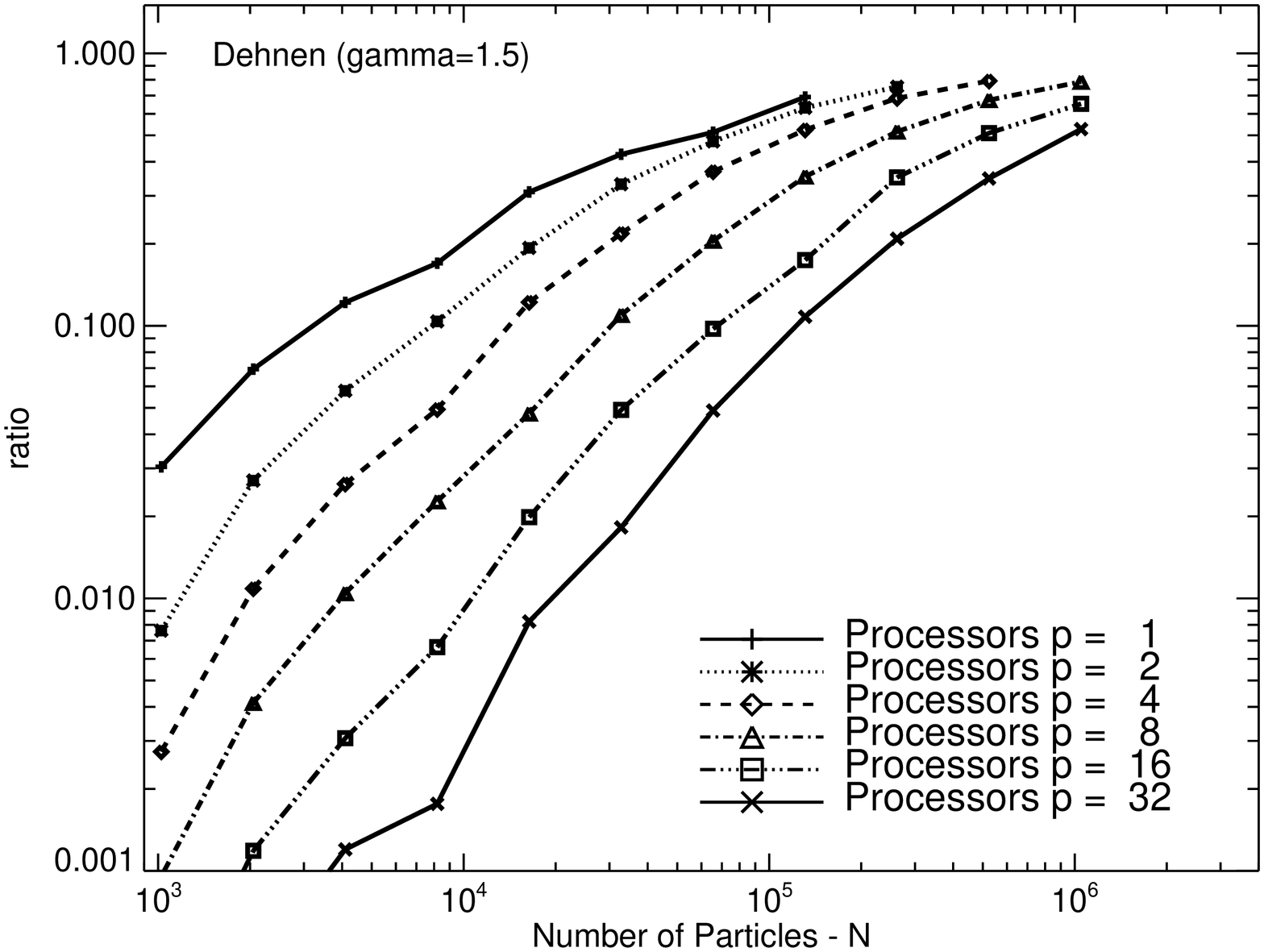} \\
   \includegraphics[width=\hsize,angle=90]{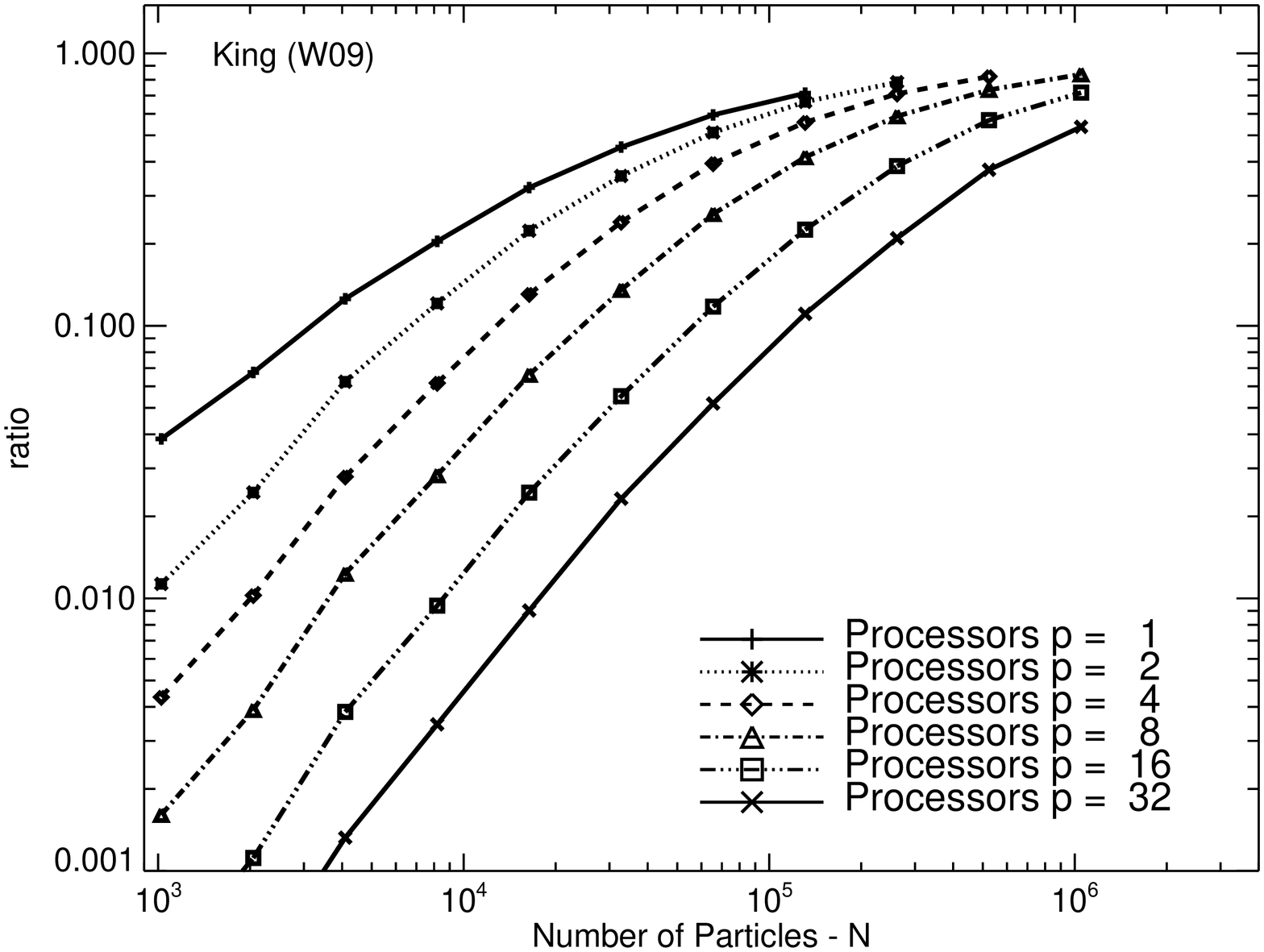} &
   \includegraphics[width=\hsize,angle=90]{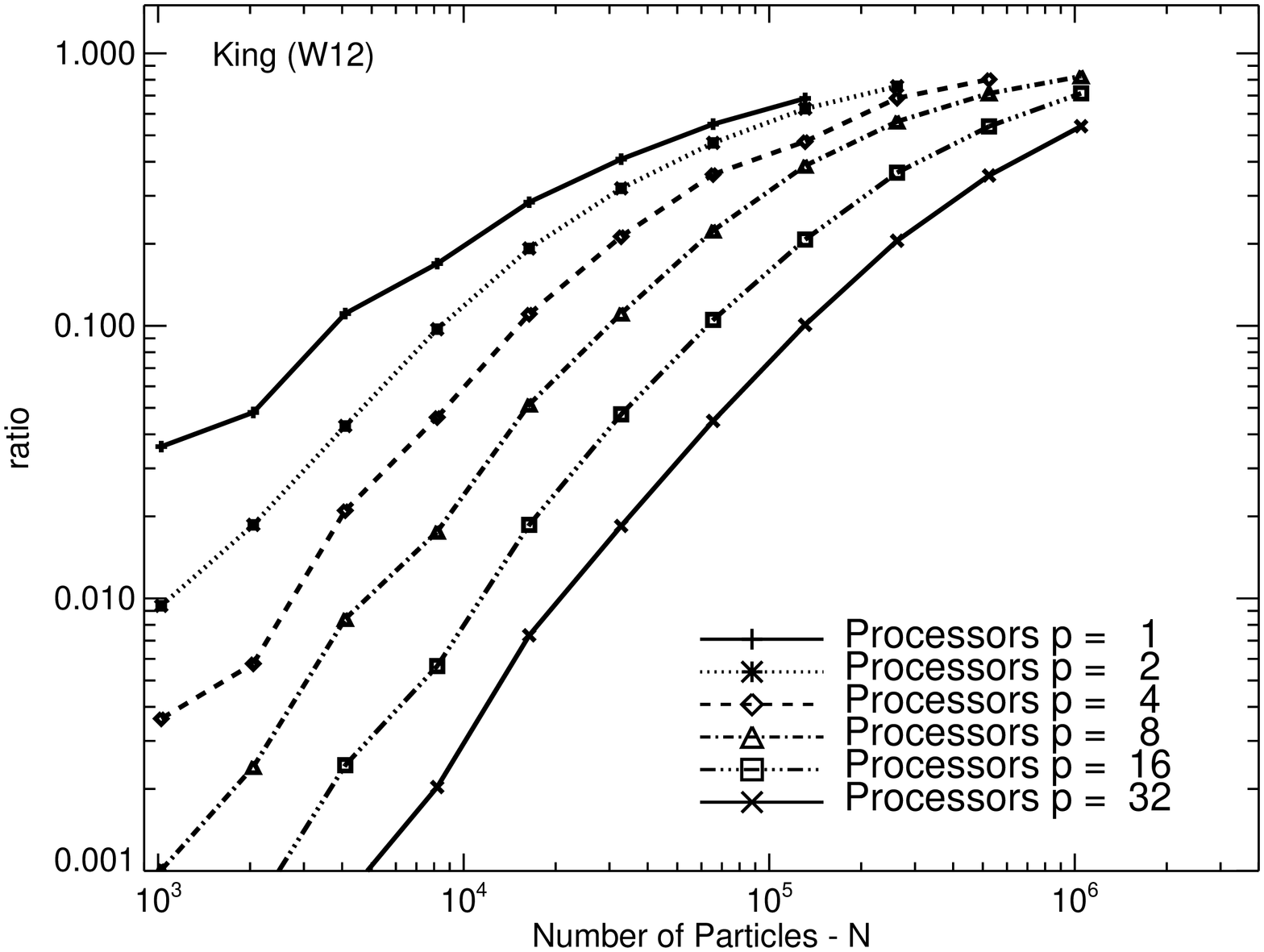} \\
   \end{tabular}} 
   \caption{Speed ratio $r$ vs. particle number $N$ for different
            numbers of processors $p$. The plots in the top row show
            the results for a Plummer model on the RIT (left) and 
            ARI (right) clusters. The remaining plots show the speed ratio
	    for Dehnen models with $\gamma=0.5$ (middle left) and
            $\gamma=1.5$ (middle right) and for King models with
            $W_0=9$ (bottom left) and  $W_0=12$ (bottom right).}  
   \label{fig_ppsr} 
\end{figure*}
%
%----------------------------------------------------------------------

We analyzed the performance of the hybrid scheme as a function
of particle number and also as a function of number of nodes;
we used $p=1, 2, 4, 8, 16,$ and $32$ nodes. 
The compute time $w$ for a total of almost
350 test runs was measured using {\tt MPI\_Wtime()}. The timing
was started after all particles had finished their initial time
step and ended when the model had been evolved for one time
unit. No data output was made during the timing interval.

Fig.~\ref{fig_ppwc} shows wallclock times $w_{N,p}$ from all
integrations on the RIT cluster as a function of particle number
$N$ and processor number $p$.  We also show results from just the
Plummer models on the ARI cluster.  For any $p$, the clock time
increases with $N$, roughly as $N^2$ for large $N$.  However when
$N$ is small, communication dominates the total clock time, and
$w$ {\it increases} with increaing number of processors.  This
behavior changes as $N$ is increased; for $N\gap 10\kk$ (the
precise value depends on the model), the clock time is found to
be a decreasing function of $p$, indicating that the total time
is dominated by force computations.  The clock time is longer for
the more centrally concentrated models since smaller time steps
are required.  As expected, the ARI cluster is faster than the
RIT cluster by about 10\% due to its newer hardware and better
communication.

The centrally concentrated King and Dehnen models tend to have
very small block sizes at small ($N\le 10^4$) particle numbers.
If such systems are integrated using the larger processor numbers,
specific features related to the hardware and software
implementation of communication (latencies) turn up, 
which are otherwise hidden by the dominating effect of large
computation and communication, e.g. bandwidth and CPU speed.
We will not discuss these effects in detail here although
their influence can be discerned in the details of
Figs. 5-9.

The speedup for selected test runs is shown in
Fig.~\ref{fig_ppsu}.  Speedup $s$ is defined as
\begin{equation}
s_{N,p} = \frac{w_{N,1}}{w_{N,p}}.
\end{equation}
The ideal speedup (optimal load distribution, zero communication and latency)
is $s_{N,p} = p$. For particles numbers $N\gap 128\kk$ the
wallclock time $w_{N,1}$ on one processor is undefined as $N$ exceeds
the memory of the GRAPE card. In that case we used $w_{N,1} =
w_{128\kk,1}\cdot(N/128\kk)^2$ assuming a simple $N^2$-scaling. 
In general, the speedup for any given particle number 
is roughly proportional to $p$ for small $p$, then reaches a maximum
before dropping at large $p$. 
The number of processors at the the point of maximum speedup is 
``optimum'' in the sense that it provides the fastest
possible integration of the given problem. 
The optimum $p$ is roughly the value at which the sum of the
communication and latency times equals the force computation
time; in the zero-latency case, $p_{opt}\propto N$ \citep{dorband-03}.
Fig.~\ref{fig_ppsu}  shows that for $N\gap 128\kk$, 
$p_{opt}\ge 32$ for all the tested models. 

Efficiency (Fig.~\ref{fig_ppeff}) $e$ is defined by 
\begin{equation}
e_{N,p} = \frac{s_{N,p}}{p}.
\end{equation}
Again, the comparison of the Plummer model on the clusters shows that
the ARI cluster performs better. On 32 nodes the efficiency can be as
high as $0.6$ for the highest $N$. Also, the efficiency does not vary
much for models that have different central
concentrations. 

As mentioned before, the theoretical peak performance of a single
GRAPE card (or node) is $f_{1,{\rm max}}\sim 131\GFlops$. We determined the compute
speed $f$ from the measured total number of force updates $N_f$
in each run. For each force update $N$ forces are calculated, i.e. the 
compute speed is
\begin{equation}
f_{N,p} = 57 \cdot \frac{N\cdot N_f}{w_{N,p}},
\end{equation}
which assumes $57$ floating point operations per force
calculation. The measured compute speed is shown in Fig.~\ref{fig_ppsp}. 
The maximum speeds reached are $2.2\TFlops$ on the RIT
cluster and $3.2\TFlops$ on the ARI cluster. 

The speed ratio $r$ is given by
\begin{equation}
r_{N,p} = \frac{f_{N,p}}{p\cdot f_{1,{\rm max}}}
\end{equation}
and shown in Fig.~\ref{fig_ppsr}. The speed ratio reached a maximum
of $\sim 0.8$ and $\sim 0.9$ on the RIT and ARI clusters
respectively. This shows again the benefits of the newer hardware 
in the ARI cluster. There are several reasons why the theoretical 
peak speed cannot be reached.  Under a realistic time step distribution,
it is impossible to keep the 48 pipelines on every GRAPE fully loaded.
Evidence for this is seen in the lower speed ratios of the models
with the more centrally-concentrated density distributions,
like the Dehnen models, in which block sizes are typically small.
In addition, the communication between the host and the GRAPE requires
a non-negligible overhead of order of 10\%.  
Communication also detracts from the performance; 
this can be seen in the slightly  better performance of the ARI 
cluster, a result of its faster network.

\section{Performance modeling}
\label{sec_pm}
In this section we present a theoretical performance model for
the execution time of a direct $N$-body code on a GRAPE cluster.  
We first consider the performance of a sequential code, which 
contains the essential elements of the performance of the GRAPE 
hardware, then consider the performance of the hybrid parallel scheme
described in $\S\,$\ref{sec:code}.

\subsection{Performance modeling of the sequential GRAPE code}

If we consider a sequential block time-step code to be used in
combination with GRAPE-6A, the time required to advance one
active particle for one integration step can be written as
\begin{equation}
T\left(1\right) = \ths\left(1\right) + \tgr\left(1\right) + \tcm\left(1\right)
\end{equation}
where $\ths\left(1\right) = \tpr\left(1\right) + \tcr\left(1\right)$ 
is the time spent on the host for the predictor and corrector operations,
$\tgr\left(1\right) = N\,\tpp$ is the time spent on the GRAPE 
for the force calculation and $\tcm\left(1\right)$ is the time spent 
in communication between the host and the GRAPE.
The communication time between GRAPE and host has three terms \citep{FMK05}:
\begin{equation}
\tcm\left(1\right) =  60\,t_i + 56\,t_f + 72\,t_j
\end{equation}
where the first term represents the time to send the predicted positions 
and velocities to the GRAPE, the second term is the time to retrieve 
acceleration, jerk, and potential from the GRAPE, and the third term
represents the time to send new data to the GRAPE memory for update.
Table \ref{tab:param} reports the parameters measured on one 
GRAPE6-A of the RIT cluster. 

\begin{figure}[ht]
\begin{center}
\includegraphics[width=7cm]{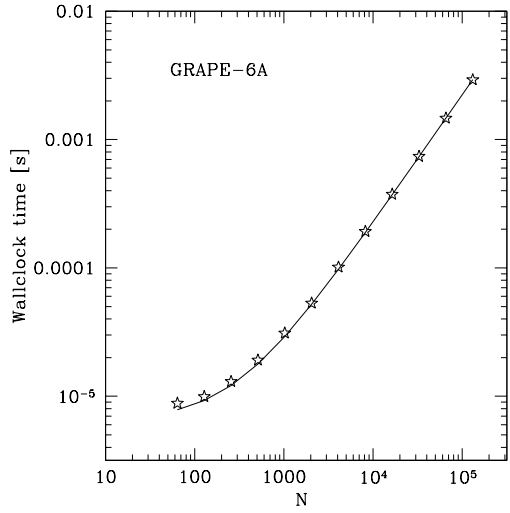}
\end{center}
\caption{Time for one particle step as a function of the number
of particles. The solid line indicates the theoretical prediction 
while the data points represent timing results on one GRAPE-6A.}
\label{fig:serial:onep}
\end{figure}

The times $\tpr$ for the predictor and $\tcr$ for the corrector
are measured on the host node.  The parameter $t_j$ is derived by
measuring the time $T_{\rm send}$ to send the data relative to
one particle to the GRAPE memory: $t_j = T_{\rm send}/72$.  We
then assume $t_i=t_f=t_j$ as in \citet{FMK05}.  The GRAPE
parameter $\tpp$ is not measured directly but derived from the
total time for the force calculation by subtracting the time for
communication between the host and the GRAPE.  This approach is
necessary since the measured time $T_{\rm force}$ for the force
calculation contains both the time for the force computation and
the communication time between host and GRAPE.  In this way
$\tpp$ is given by $\tpp = \left(T_{\rm force} - \left(60\,t_i +
56\,t_f\right)\right)/N$.  The total wallclock time for one
particle step is shown in Fig.\,\ref{fig:serial:onep} for
different particle numbers, and compared with timing data on one
GRAPE-6A.
The agreement between the model and the data is very good for
particle numbers larger than about 1K. For smaller $N$, there is
a small deviation of the model from the data. The deviation is
due to a slightly different value of $\tpp$ when the GRAPE memory
holds a very small number of particles. Given the fact that we
are not interested in such small particle numbers, we ignore this
effect and consider a fixed $\tpp$ throughout our analysis.

\begin{table}
  \caption{Performance parameters of one GRAPE-6A board}
  \label{tab:param}
  \begin{center}
    \begin{tabular}{cccc}
      \hline
      \hline
      $\tpr\left(1\right)$ & $\tcr\left(1\right)$ & $\tpp$ & $t_j$\\
      \hline
       $(1.1\pm0.2)\times10^{-7}$ & $(3.4\pm0.3)\times10^{-7}$ 
      & $(2.2\pm0.5)\times10^{-8}$ & $(3.1\pm0.5)\times10^{-8}$ \\ 
      \hline
      \hline
    \end{tabular}
  \end{center}
\end{table}
\bigskip

The total time required to advance a block of active particles 
of size $s$ can then be written as
\begin{equation}
T\left(s\right) = \ths\left(s\right) + \tgr\left(s\right) + \tcm\left(s\right)
\end{equation}
where 
\begin{eqnarray}
\ths\left(s\right) &=& \tpr\left(s\right) + \tcr\left(s\right) = \ths\left(1\right)\, s\, , \nonumber \\
%\end{equation}
%\begin{equation}
\tgr\left(s\right) &=& N\,  \tpp \left[s/48\right]\, , \nonumber \\
%\end{equation}
%\begin{equation}
\tcm\left(s\right) &=& 60\, t_i s + 56\, t_f s + 72\, t_j s = 188\, t_j s\,,
\end{eqnarray}
with $[x/y] = (\rm int) (x/y) +1$.
Since the GRAPE pipeline accepts a maximum of 48 particles,
$\tgr$ is a step function with a step in correspondence
of multiples of 48 in block size.

\begin{figure}
\begin{center}
\includegraphics[width=6.2cm]{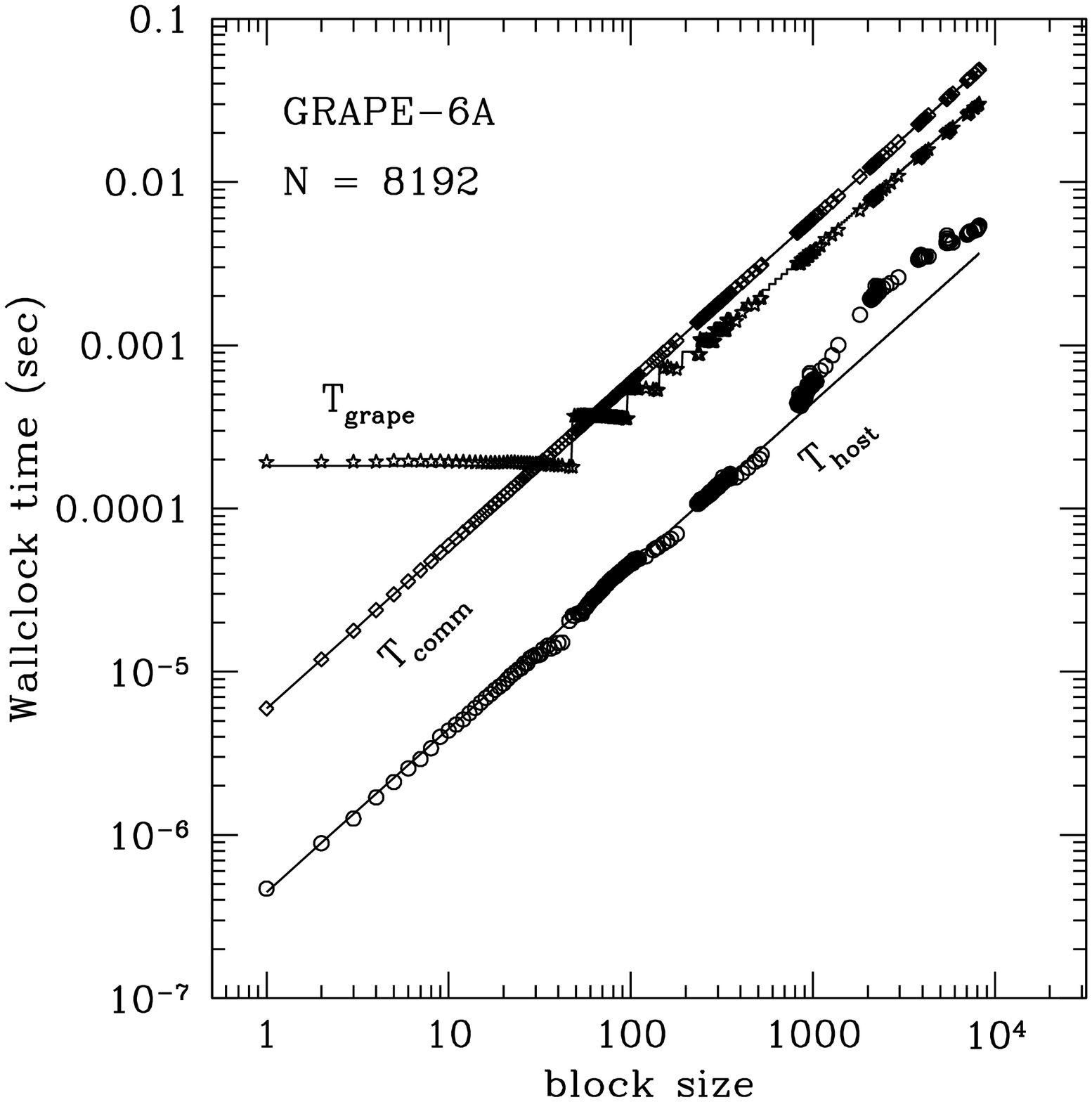}
\includegraphics[width=6.2cm]{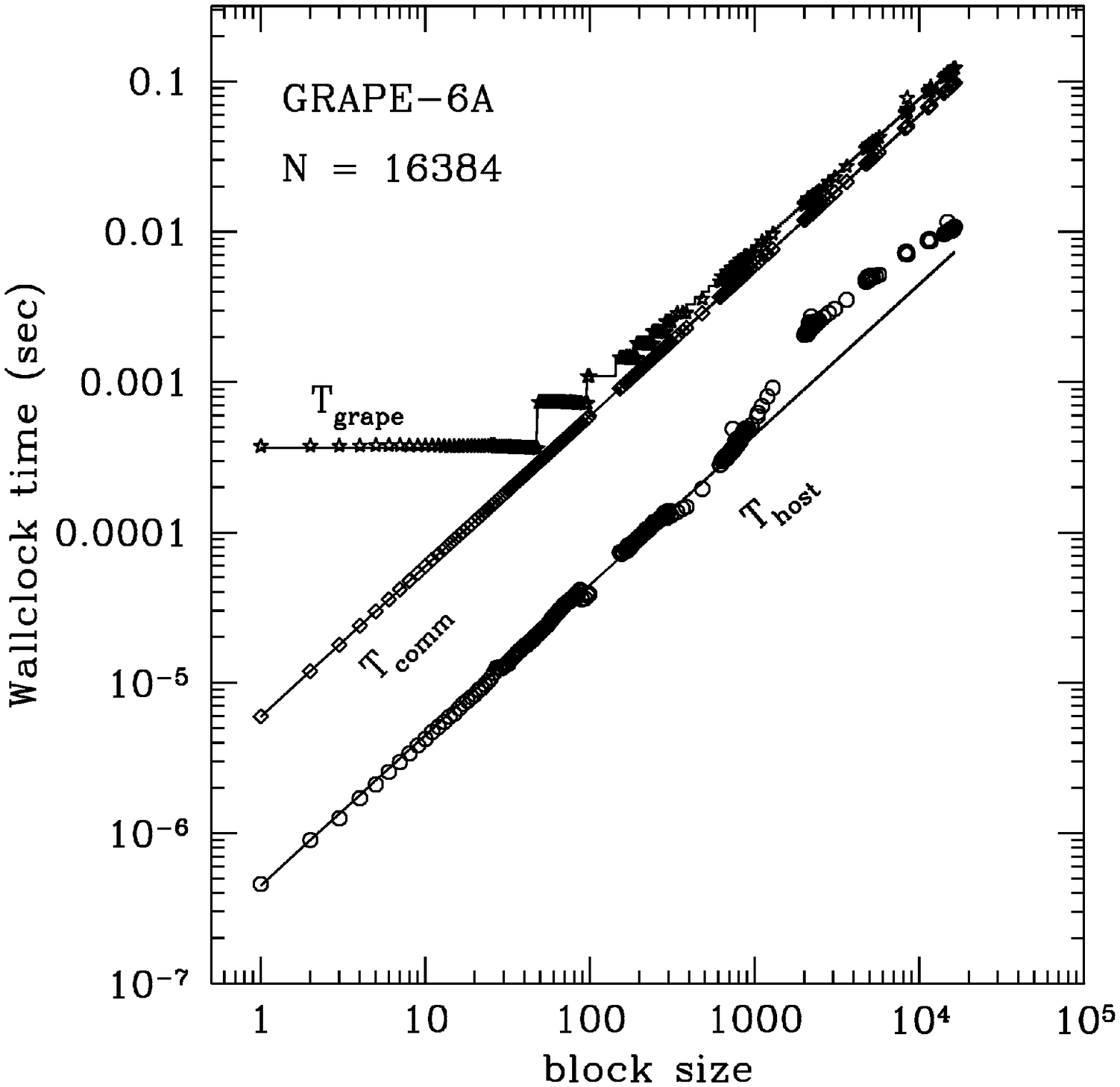}
\includegraphics[width=6.2cm]{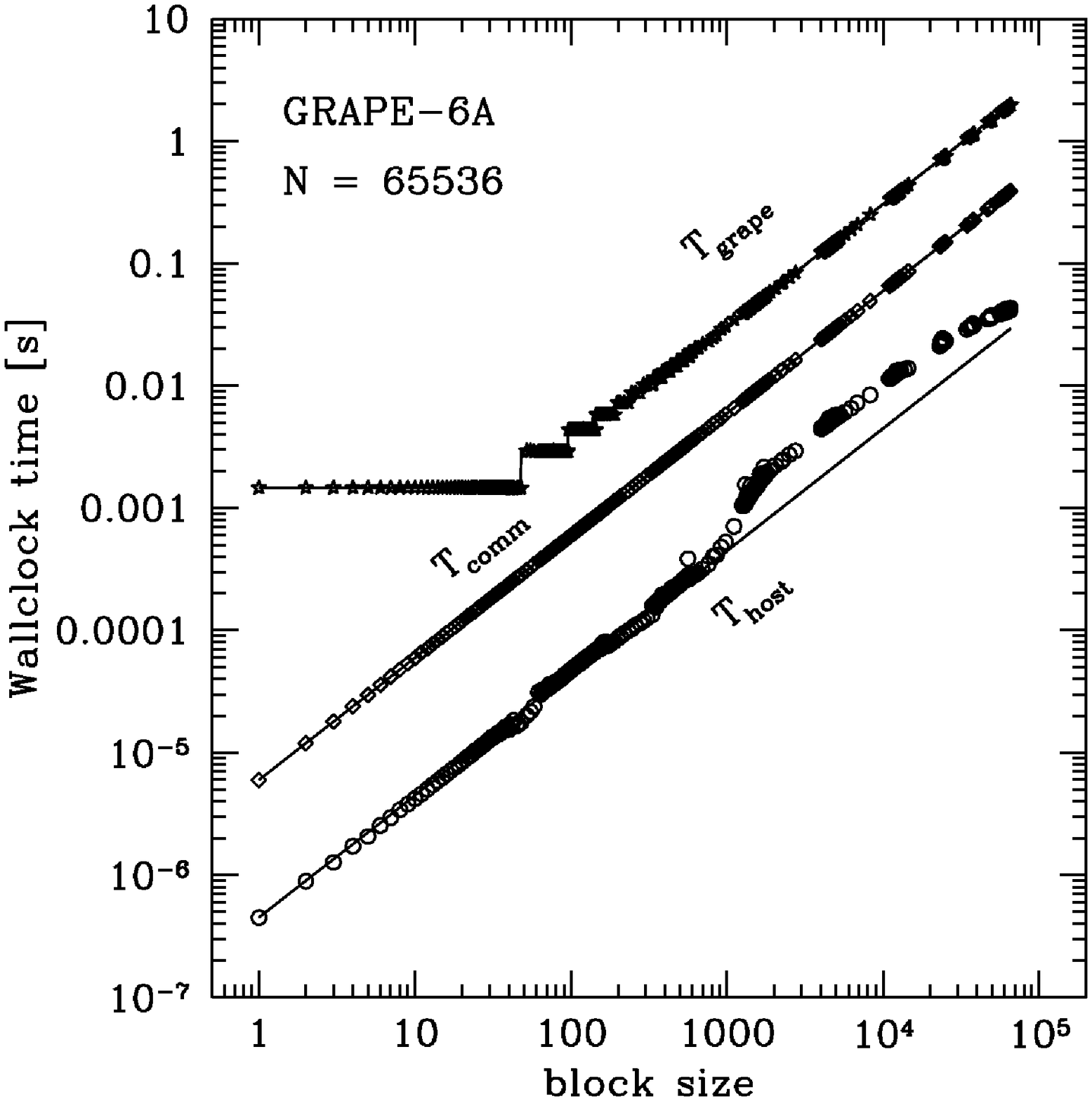}
\includegraphics[width=6.2cm]{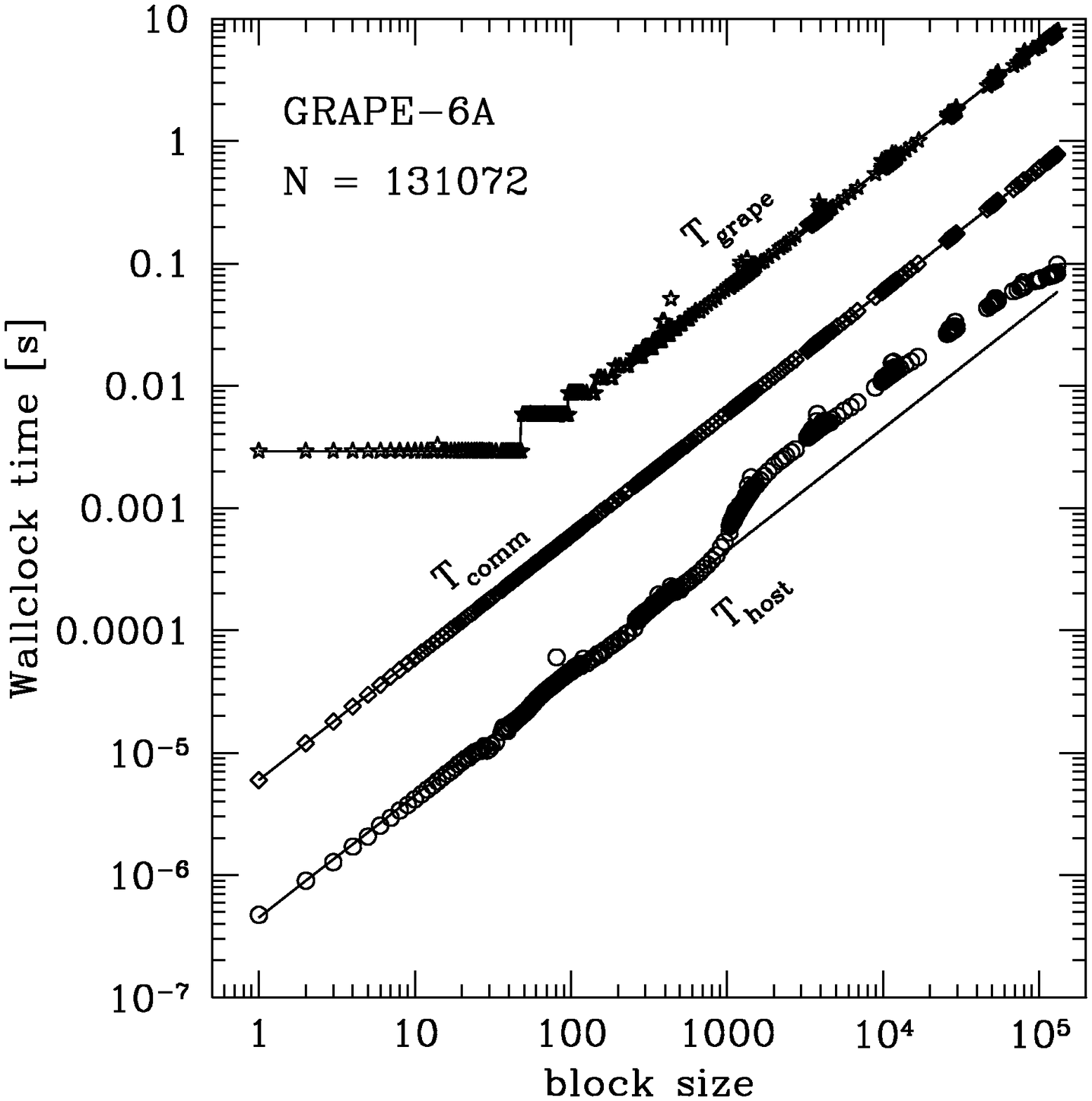}
\end{center}
\caption{Times, predicted by the theoretical model, spent on the host, 
the GRAPE and in communication as a function of the block size 
for a sequential GRAPE code with block time steps. 
Data points are actual timing 
measurements on one GRAPE-6A of the RIT cluster.}
\label{fig:serial:step}
\end{figure}

Fig.\,\ref{fig:serial:step} shows a comparison between the 
predicted times (solid lines) for given block sizes $s$ 
and timing measurements (data points) conducted on the RIT cluster.
The different plots refer to Plummer models with $N=8\kk, 16\kk,
64\kk, 128\kk$.  Given the errors on the timing measurements and
on the parameters reported in Table\,\ref{tab:param}, there is
good agreement between the data and the performance model.  For
the $N=8\kk$ model, the time spent in communication between the
host and the GRAPE is larger then the time spent on the GRAPE
itself for the force calculation. For $N\gtrsim 16\kk$, the time
for the force calculation becomes larger then the communication
time and for the $N=128\kk$ system the total execution time is
dominated by the force calculation on the GRAPE.  This shows that
the use of GRAPE hardware for $N$-body
simulations\index{N-body!simulations} is most efficient in the
case of large systems, which are the most interesting from the
scientific point of view.  The non-linear increase in the host
time for block sizes larger than about 1000 is likely due to
cache misses.

\begin{figure}[ht]
\begin{center}
\includegraphics[width=8cm]{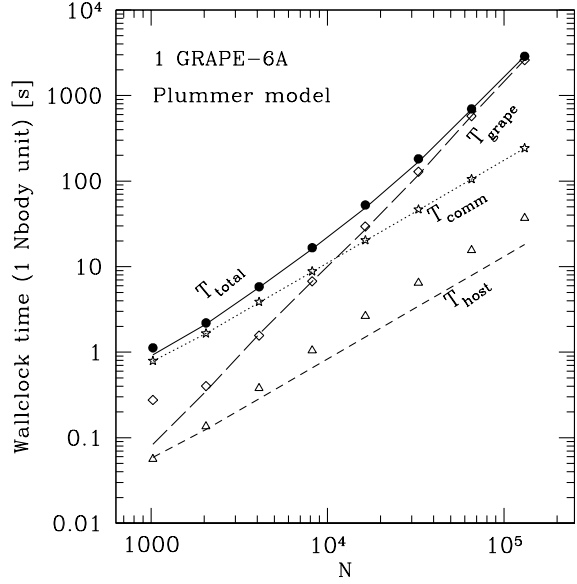}
\end{center}
\caption{Comparison between the predictions by the theoretical
model and the timing measurements for the integration
of Plummer models over one $N$-body unit.
The long-dashed line indicates the time spent on the GRAPE 
for the force calculation, the dotted line indicates the time 
spent in communication between the host and the GRAPE, 
the dashed line indicates the time spent on the host 
and the solid line represents the total execution time given 
by the sum of the three separate times.
The data points indicate the results of the timing experiments
conducted on a GRAPE-6A of the RIT cluster.}
\label{fig:serial:nbody}
\end{figure}

The prediction and the measurements are independent of the chosen
$N$-body model as long as the execution time is expressed as a
function of the block size.  In order to predict the execution
time for the integration of a system over one $N$-body unit (or
any other physical time), it is necessary to know the block size
distribution for the model under consideration.  In the case of a
Plummer model with given $N$, the total execution time over one
$N$-body time unit can be estimated by considering the average
value of the block size $\langle s \rangle$ and the total number
of integration steps $n_{\rm steps}$ in one $N$-body unit,
\begin{equation}
\label{eq:tsnb}
T_N = T_N(\langle s \rangle)\,n_{\rm steps}\,.
\end{equation}
We have measured the average block size and the number of steps
in one $N$-body unit for Plummer models of different $N$ and applied
Eq.\,\ref{eq:tsnb} to the prediction of the total execution times
for the same models.
Fig.\,\ref{fig:serial:nbody} shows a comparison betweeen the 
predicted execution time for the integration of a Plummer model
over one $N$-body unit and timing measurements on a single GRAPE-6A.
The model satisfactorily predicts the time spent on the host,
on the GRAPE and in communication for particle numbers 
$N\gtrsim 2\kk$.
For smaller $N$, deviations from the prescription given
by Eq.\,\ref{eq:tsnb} are more likely to occur and to affect
the modeling. In particular, the block size is generally small
and the average is generally not a good representation 
of the global behavior of the system.

\subsection{Performance modeling of the parallel GRAPE code}
In the case of the hybrid scheme, the total execution time
can be written as
\begin{equation}
  T = \ths + \tgr + \tcm + \tmpi
\end{equation}
where 
%$\ths$ represents the time spent on the host computer, 
%$\tgr$ represents the time spent on the GRAPE, 
%$\tcm$ denotes the communication time between the GRAPE
%and the host and 
$\tmpi$ indicates the time spent in communication among the
nodes.  If $s$ is the block size at a specific step during the
integration and $s_{\rm max} = \max_{i=1,\dots p} \left\{s_i
\right\}$ is the maximum of the local blocks on the different
nodes, the time spent on the host is given by
\begin{equation}
\ths = \tpr\left(s_{\rm max}\right) + \tcr\left(s_{\rm max}\right) = 
\tpr\left(1\right)\,s_{\rm max} + \tcr\left(1\right)\,s_{\rm max}\, ,
\end{equation}
the time spent on the GRAPE is given by
\begin{equation}
  \tgr = \frac{N}{p}\,\tpp \left[s/48\right]\, ,
\end{equation}
the time spent in communication between the host and the GRAPE is
\begin{equation}
\tcm = 60\, t_i s + 56\, t_f s + 72\, t_j\,s_{\rm max}\,,
\end{equation}
and the time spent in communication among the nodes is given by
the sum of the time spent in each MPI call.
The time $\tmpi$ is dominated by two calls to the function 
{\tt  MPI\_Allreduce} 
and three calls to the function {\tt  MPI\_Allgatherv}.
We adopt the following models for the MPI functions:
\begin{eqnarray}
T_{\tt MPI\_Allgatherv} &=& \left( \alpha + \beta\,x\right) log_2 p ,
\nonumber \\
T_{\tt MPI\_Allreduce} &=& \left( \delta + \gamma\,x\right) log_2 p
\end{eqnarray}
where $x$ represents the size of transferred data measured in bytes
and $\alpha$, $\beta$, $\delta$, $\gamma$ are parameters
obtained by fitting timing measurements on the RIT cluster 
(see Table\,\ref{tab:mpi}).

\begin{table}[h]
  \caption{Fit parameters for modeling of the MPI functions.}
  \label{tab:mpi}
  \begin{center}
    \begin{tabular}{cccc}
      \hline
      $\alpha$\,[sec] & $\beta$\,[sec] & $\delta$\,[sec] & $\gamma$\,[sec]\\
      \hline
      $1.2\times10^{-5}$ &  $2.5\times10^{-9}$ & $1.0\times10^{-5}$ &  $1.0\times10^{-8}$\\
      \hline      
    \end{tabular}
  \end{center}
\end{table}

\begin{figure}[ht]
\begin{center}
\includegraphics[width=6.2cm]{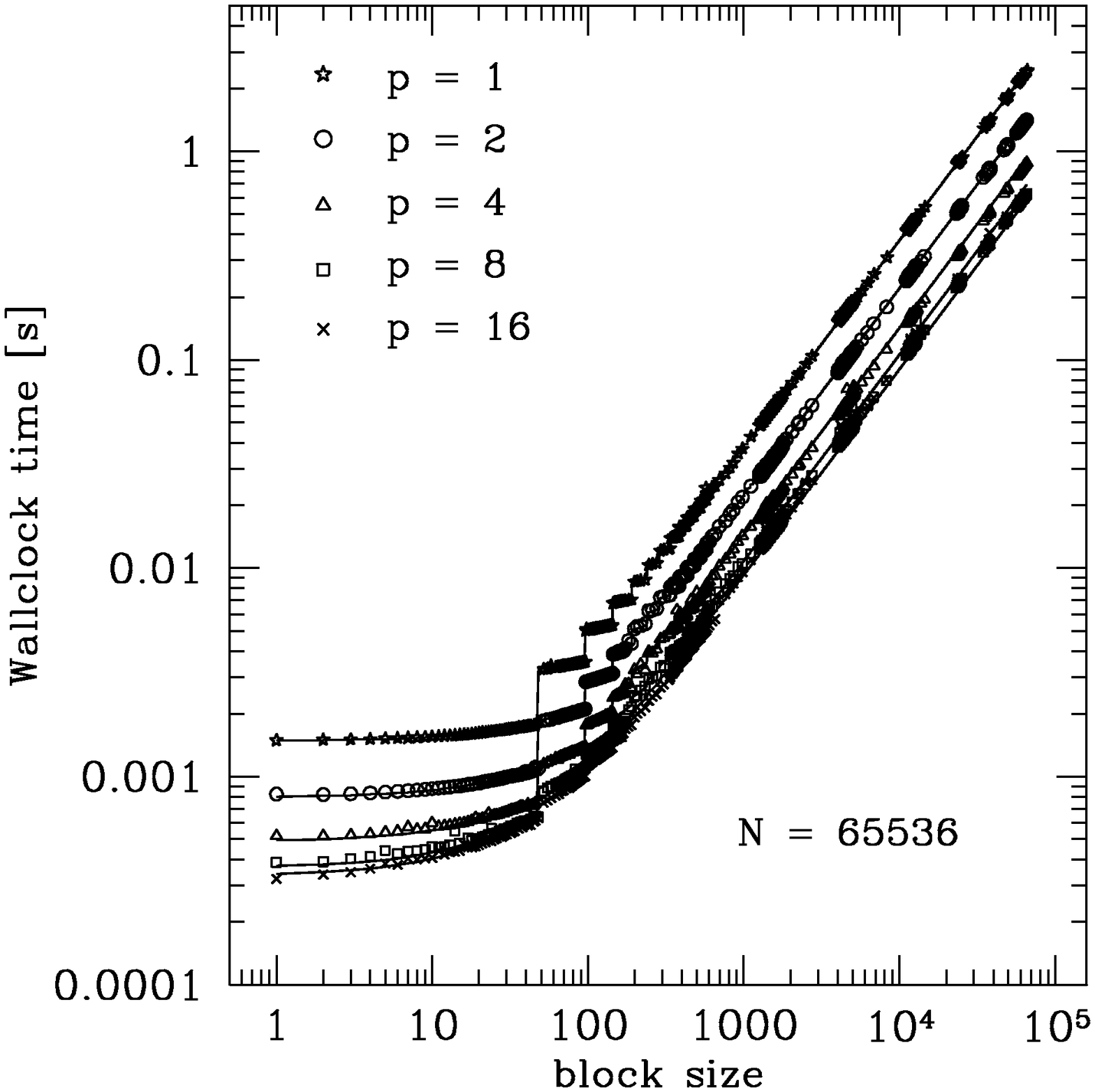}
\includegraphics[width=6.2cm]{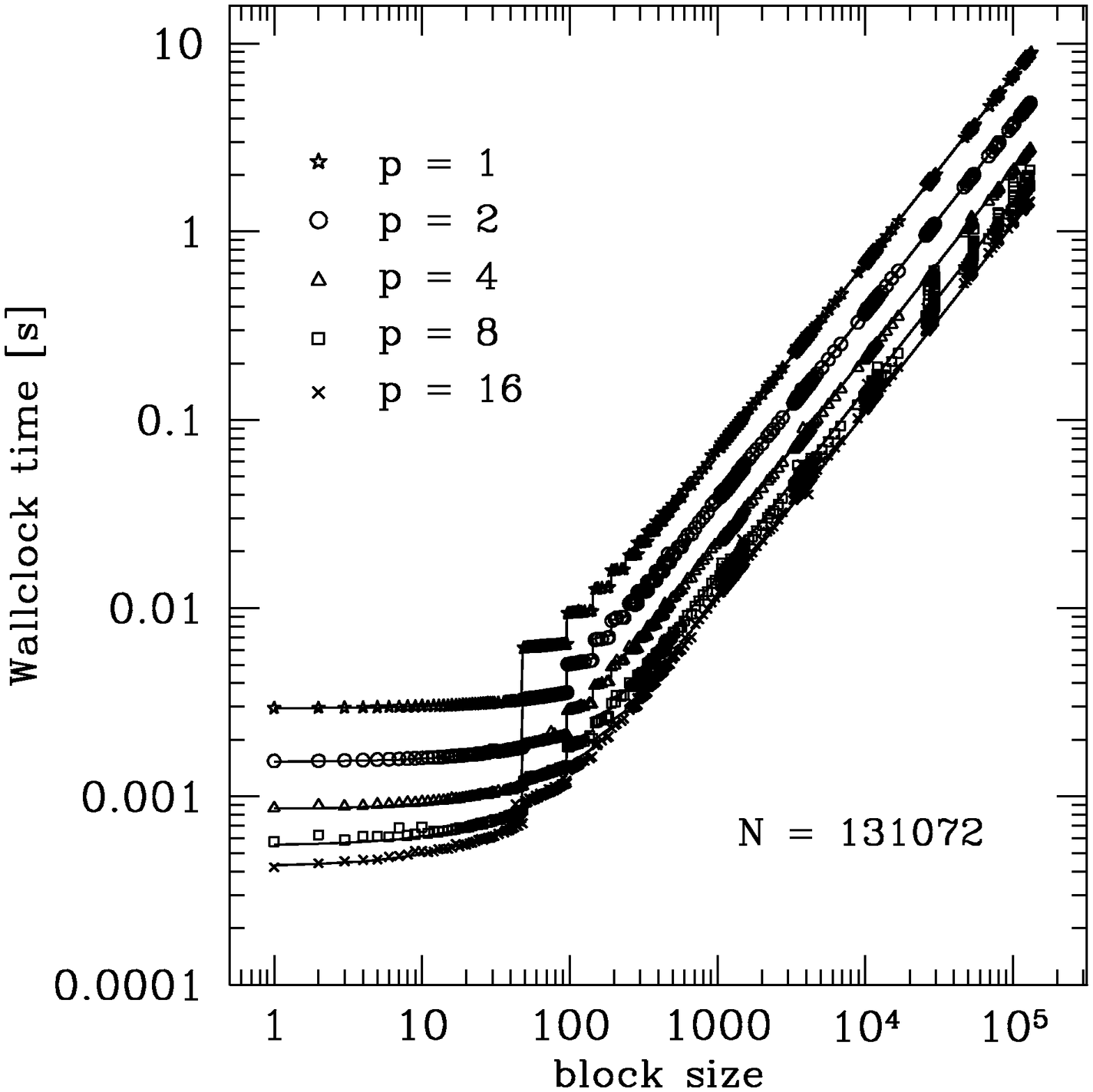}
\end{center}
\caption{Comparison between the execution time predicted by the model
for the integration of a Plummer model with $N$=65536 (left) 
and $N$ = 131072 (right) and timing experiments on the RIT cluster. 
The total execution time is plotted as a function of the block size.}
\label{fig:parallel:step}
\end{figure}

Fig.\,\ref{fig:parallel:step} reports the comparison between the
prediction for the total execution time as a function of the
block size at one specific step and the timing results on the RIT
cluster.  As in the case of the theoretical model, the times
spent on the host, the GRAPE, in communication with the GRAPE and
in communication among the nodes are measured separately and then
added together.

In order to estimate the total execution time of the parallel scheme
over one $N$-body unit, we consider the average value of the block size 
$\langle s \rangle$ and the total number of integration steps 
$n_{\rm steps}$ in one $N$-body unit for Plummer models 
with different particle numbers.
Similarly to the case of the sequential code, we can approximate
the total execution time for a particle number $N$ and processor
number $p$ as
\begin{equation}
\label{eq:tpnb}
T_{N,p} = T_{N,p}(\langle s \rangle)~n_{\rm steps}\,.
\end{equation}

\begin{figure}[ht]
\begin{center}
\includegraphics[width=8cm]{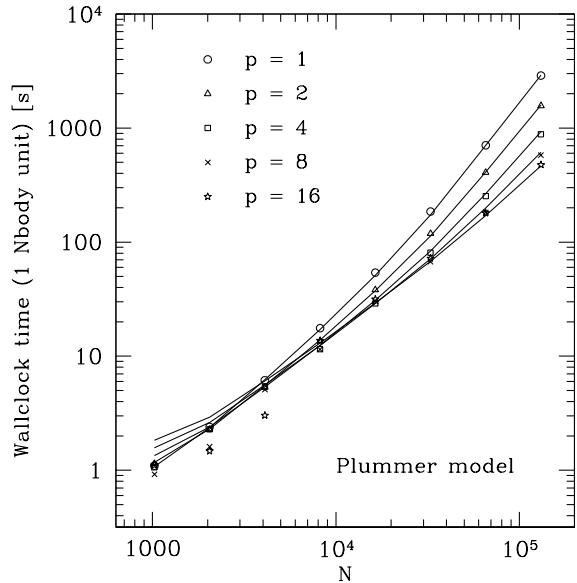}
\end{center}
\caption{Comparison between the predictions by the theoretical
model (solid lines) and the timing measurements (data points) 
for the integration of Plummer models over one $N$-body unit.
The different lines refer to different processor numbers.}
\label{fig:parallel:nbody}
\end{figure}

Fig.\,\ref{fig:parallel:nbody} shows a comparison betweeen the 
predicted execution time for the integration of different Plummer models
over one $N$-body unit and timing measurements.
For the theoretical prediction we have used Eq.\,\ref{eq:tpnb}
and measured values for the average block size and the number of
steps in one $N$-body unit.  The agreement between the model and
the data is good for large particle numbers while deviations
appear for small $N$.

\section{Discussion}

The performance of a direct-summation, parallel $N$-body code 
has been evaluated on two new computer clusters incorporating
GRAPE special-purpose accelerator boards.
Parallelization was carried out using a hybrid method incorporating
aspects of both systolic (``ring'') and broadcast (``copy'')
algorithms, in order to make the most efficient use of the GRAPE pipelines. 
The algorithm exhibits asymptotically an $\mathcal{O}(Np)$ scaling in 
communication complexity and an $\mathcal{O}(N^2/p)$ scaling in 
computation time.
Benchmark simulations were carried out using a set of galaxy models 
having realistically high degrees of central concentration.
Using one million particles and 32 nodes, the clusters achieved
a sustained performance of 50\% -- 100\% of the theoretical peak 
($\sim 4\TFlops$).
When run on general-purpose parallel computers, $N$-body codes
like ours typically only achieve a few per cent of peak performance;
hence, our special-purpose computer clusters are competitive
with the fastest computers in the world when applied to the
gravitational $N$-body problem.

We presented a simple model that predicts the performance of the
$N$-body code as a function of processor number $p$, particle
number $N$, and hardware constants. 
The model reproduces the observed performance very well,
and can be used to predict the performance of the code
on clusters with different hardware characteristics.
The model supports our finding that the performance depends
critically on having very fast and low-latency communication hardware.

It is convenient to discuss the performance of the parallel 
$N$-body code in terms of a working point $P$
defined by a pair of values $P=(p,N)$, where $P$ is 
given by the condition that the time required for communication and
computation be approximately equal.
When using all 32 of the processors on our clusters, 
this point is reached for $N\approx 10^6$. 
For larger $N$, the system operates at or near optimal
speedup and efficiency (see Figs. 6 and 7). 
At a given $p$, the linear scaling of total computation time with
$N$ at low $N$ (communication-dominated) changes to the asymptotic, 
$N^2$ scaling (computation dominated) roughly at $P$.
Increasing $N$ beyond its value at $P$ would still achieve near-optimal
speedups, but there would be room to increase $p$
without sacrificing efficiency (if more nodes were available).
Increasing $p$ beyond $P$ at fixed $N$ is not useful
because communication overhead will start to dominate, 
leaving the GRAPEs idle for part of the time.
The design goal of our clusters was to
simulate $\sim 10^6$ particles at high efficiencies,
and we have shown by our performance tests that this goal has been 
achieved.

\begin{figure}[ht]
\begin{center}
\includegraphics[width=10cm,angle=-90.]{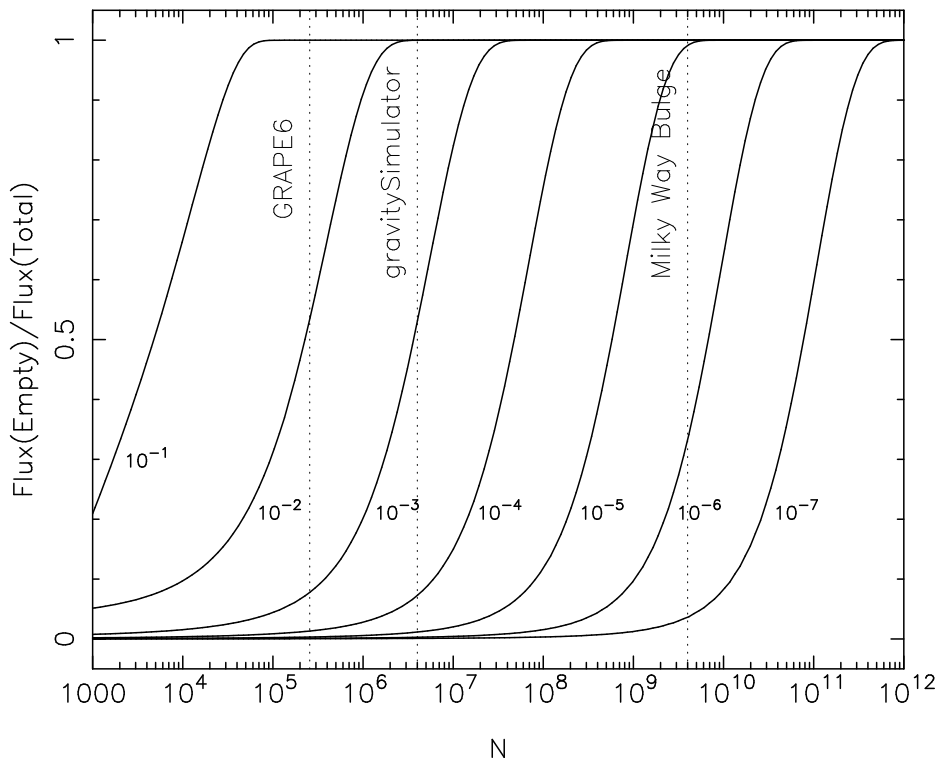}
\end{center}
\caption{This figure demonstrates what particle
numbers are needed to achieve the separation
of time scales discussed
in the text, when simulating nuclei containing a 
central sink, e.g. a single or binary BH.
Vertical axis is the fraction of the flux 
into the ``sink'' originating from orbits that
are in the empty-loss-cone regime; in real
galaxies, this ratio is close to unity.
Curves are labelled by $r_{LC}/r_h$, where $r_{LC}$
is the linear size of the central ``sink''
and $r_h$ is the BH influence radius.
For a binary SBH, $0.1\le r_{LC}/r_h\le 10^{-3}$.
}
\label{fig:empty}
\end{figure}

How large an $N$ is required to accurately
represent the evolution of a dense stellar system like a
galactic nucleus?
A minimum condition is that $N$ be large enough to enforce
a strict separation of time scales between the period
$T_{orb}$ of an orbit and the relaxation time $T_{ref}$; the 
latter is the time for two-body
gravitational scattering to randomize
velocities \citep{spitzer-87}.
In real galaxies, $T_{rel}\gg T_{orb}$, i.e. the integrity
of stellar orbits is maintained for many orbital periods.
In an $N$-body simulation, one has
\begin{equation}
T_{rel}\approx {0.1 N\over \ln N} T_{orb} 
\label{eq:tr}
\end{equation}
\citep{aarseth-03}, and the separation of time scales 
can be achieved with modest particle numbers, $N\gap 10^3$.
Relaxation-driven processes like core collapse
can be simulated using $N$-body codes and the numerical evolution
rates scaled to those in real galaxies using an equation like (\ref{eq:tr})
\citep{SA-96,SMK-05}.

In simulations of galactic nuclei, a much stronger separation of time scales
must be achieved if the results are to be scaled to real systems,
implying larger values of $N$.
Galactic nuclei contain ``sinks,'' regions near the
center where stars are lost or captured.
For instance, the supermassive black hole (BH) at the
center of the Milky Way galaxy disrupts stars that pass within a 
distance $r_{tidal}\approx 10^{-6}$ pc of it.
Another example is a galaxy containing a binary supermassive BH:
stars that pass within a distance $\sim a$ of either BH
are ejected by the gravitational slingshot,
where $10^{-3}\ {\rm pc} \lap a \lap 10^{-1}$ pc 
is the semi-major axis of the binary.
In this case, the radius of the ``sink'' is $\sim a \gg r_{tidal}$.
Single or binary supermassive BHs are believed to be
generic components of galaxies \citep{FF05,living},
and the structure and evolution of nuclei may be shown to
depend critically on the rate at which stars are lost 
to the central sink \citep{merritt-06}.

In real galaxies, relaxation times are long enough that most stars 
would diffuse gradually 
-- i.e., over many orbital periods --
onto so-called loss-cone orbits that intersect the capture 
sphere of radius $r_{tidal}$ or $a$.
This means that the loss cone is essentially empty, since
the removal time is much shorter than the diffusion time.
In order to reproduce a diffusive loss-cone repopulation
in an $N$-body simulation, the relaxation time must be large
enough that
\begin{equation}
T_{rel} > \left({r\over r_{LC}}\right) T_{orb} \gg T_{orb}
\label{eq:loss}
\end{equation}
\citep{LS-77}.
Here, $r_{LC}$ is the radius of the disruption ($\sim r_{tidal}$) 
or ejection ($\sim a$) sphere at the center of the galaxy, 
and $r$ is the typical size of a stellar orbit.
Equation~(\ref{eq:loss}) reflects the fact that
scattering into a loss-cone orbit occurs in much less than
one relaxation time, hence maintaining an empty loss
cone requires that relaxation times be much longer than
orbital periods.
This is a stricter condition than $T_{rel}\gg T_{orb}$ and
implies a larger $N$.

Fig.~\ref{fig:empty} presents a more careful analysis
along these lines.
Shown there is the fraction of the flux into the
central sink originating from orbits that are in the
empty loss cone regime, as a function of $N$ and $r_{LC}$;
the latter is expressed in units of $r_h$, 
the radius of gravitational influence
of the central (single or binary) BH of mass $M_\bullet$.
(The influence radius is the radius containing a mass
in stars equal to twice $M_\bullet$.)
In the case of a binary supermassive BH,
$r_{LC}/r_h\approx a/r_h$ is roughly $10^{-1}$
at the time of binary formation, falling
to $\sim 10^{-3}$ when the two BHs are close enough
to coalesce.
Fig.~\ref{fig:empty} suggests that particle numbers
accessible to a single GRAPE-6 ($\sim 0.25$M) can 
only reproduce the empty loss cone regime characteristic
of real galaxies during the early phases of
binary evolution.
The values of $N$ achievable on gravitySimulator ($\sim 4M$)
allow the evolution of a binary to be followed
to separations $\sim 1/10$ as small.
These estimates are consistent with the results of
recent simulations which reproduce
diffusion-limited evolution of massive binaries
with $N\approx 10^6$ \citep{MF04,BMS05}.
Simulating loss cone dynamics around the much smaller
tidal disruption sphere of a single BH ($r_{LC}\approx 10^{-6}r_h$)
would require considerably larger $N$, well beyond the
capabilities of existing or planned computers.
However the dynamics in this regime could be qualitatively
reproduced by adjusting the size of the capture sphere in an $N$-body
simulation such that the bulk of the stars scattered into
the BH are on orbits respecting the correct ratio of $T_{rel}$
to $T_{orb}$.

The $T_{rel}\sim N$ scaling of Eq.~(\ref{eq:tr}) implies
an effectively $\sim N^3$ scaling for calculations that extend 
over one relaxation time of the system.
This scaling makes it very expensive to simulate the 
``collisional'' evolution of stellar systems using
the full, $4\times 10^6$ particle number allowed by 
the combined GRAPE memories on the RIT and ARI clusters.
One way to accelerate the computations without significant
loss of accuracy would be to use the Ahmad-Cohen (AC)
neighbor scheme \citep{AC73}, 
as implemented in codes like NBODY5 or NBODY6 \citep{aarseth-99}.
In the AC scheme, the full forces are computed only 
every tenth timestep or so; in the smaller intervals,
the forces from nearby particles (the ``irregular''
force) are updated using a high order scheme, while those from 
the more distant ones (the ``regular'' force) are extrapolated using a 
second-degree polynomial. 
A parallel implementation of NBODY6, including the AC
scheme, exists, but only for general-purpose parallel computers 
\citep{spurzem-99,khalisi-03}; the algorithm has not yet been
adapted to systems with GRAPE hardware.

Greater speedups could also be achieved by increasing the number
$p$ of processors, but only if communication costs are held low.
One way to do this is via a variant of the \cite{lippert-98a,lippert-98b}
hyper-systolic (HS) algorithm.
In its basic form, the HS algorithm reduces the number of
data transfers by storing the shifted data on each node.
The number of shifts, and hence the communication time, 
is reduced from $\mathcal{O}(p)$ to $\mathcal{O}(\sqrt{p})$ and the
memory requirements are increased by a similar factor.
The HS speedup is smaller when only a subset of the full
$N$ particles are advanced at every time step however
\citep{dorband-03}.
\citet{makino-02} has presented a direct $N$-body summation code optimized 
for a quadratic layout of processor ($p$ required to be a square number), which 
is in fact a simplified version of the hypersystolic algorithm proposed by 
\citet{lippert-98b}. In Makino's case the asymptotic
scaling of communication is $\mathcal{O}(N/\sqrt{p})$ (for calculation
cost there is no difference to our code), which would allow to use
much larger numbers of processors.  However, performance tests of a
HS $N$-body algorithm on actual hardware have apparently not been
carried out.

Implementation of the AC and/or HS schemes should permit
effective use of the RIT and ARI clusters with the full
particle numbers permitted by the combined GRAPE memories.
Still larger $N$ could be attained by implementing a
hybrid scheme, e.g. coupling a direct-summation algorithm
with a tree \citep{MA-93} or basis-function representation
\citep{HSS-02} for the distant particles,
or a hierarchical generalization of the Ahmad-Cohen
neighbor scheme.

%\section*{Acknowledgements}
%\begin{acknowledgments}
\ack
SH and DM were supported by grants 
AST-0420920 and AST-0437519 from the NSF, grant NNG04GJ48G from NASA,
and grant HST-AR-09519.01-A from STScI.
AG and SPZ acknowledge support from the Netherlands Organization for
Scientific Research (NWO, \#635.000.001), the Royal Netherlands
Academy of Arts and Sciences (KNAW) and the Netherlands Research
School for Astronomy (NOVA). RS acknowledges support from the
Volkswagen Foundation under grant No. I80 041-043, the Ministry of
Science, Education and Arts of the state of Baden-W\"urttemberg,
Germany, and the German Science Foundation under SFB 439 at the
University of Heidelberg. 
PB acknowledges
support from the German Science Foundation under SFB 439 and INTAS
grant IA-03-59-11. 
This research was supported in part by the National
Science Foundation under Grant No. PHY99-07949.
Some of the $N$-body calculations presented here were carried out at
the Center for the Advancement of the Study of 
Cyberinfrastructure at RIT whose support is gratefully
acknowledged.
%\end{acknowledgments}

%\appendix
%\section{Appendix material}

\bibliographystyle{elsart-harv}         
\bibliography{harfst}

\end{document}